\documentclass{pasj00}
\draft

\begin{document}
\SetRunningHead{Author(s) in page-head}{Running Head}
\Received{2000/12/31}%{yyyy/mm/dd}
\Accepted{2001/01/01}%{yyyy/mm/dd}

\title{A Comprehensive Study of Short Bursts from SGR\,1806$-$20 and SGR\,1900$+$14 Detected by HETE-2}

%%% begin:list of authors
\author{
  Yujin E. \textsc{Nakagawa}\altaffilmark{1}
  Atsumasa \textsc{Yoshida}\altaffilmark{1,2}
  Kevin \textsc{Hurley}\altaffilmark{3}
  Jean-Luc \textsc{Atteia}\altaffilmark{4}\\
  Miki \textsc{Maetou}\altaffilmark{1}
  Toru \textsc{Tamagawa}\altaffilmark{2}
  Motoko \textsc{Suzuki}\altaffilmark{2}
  Tohru \textsc{Yamazaki}\altaffilmark{1}\\
  Kaoru \textsc{Tanaka}\altaffilmark{1}
  Nobuyuki \textsc{Kawai}\altaffilmark{2,5}
  Yuji \textsc{Shirasaki}\altaffilmark{2,6}
  Alexandre \textsc{Pelangeon}\altaffilmark{4}\\
  Masaru \textsc{Matsuoka}\altaffilmark{7}
  Roland \textsc{Vanderspek}\altaffilmark{8}
  Geoff B. \textsc{Crew}\altaffilmark{8}
  Joel S. \textsc{Villasenor}\altaffilmark{8}\\
  Rie \textsc{Sato}\altaffilmark{5}
  Satoshi \textsc{Sugita}\altaffilmark{1}
  Jun'ichi \textsc{Kotoku}\altaffilmark{5}
  Makoto \textsc{Arimoto}\altaffilmark{5}\\
  Graziella \textsc{Pizzichini}\altaffilmark{9}
  John P. \textsc{Doty}\altaffilmark{8,10}
  and
  George R. \textsc{Ricker}\altaffilmark{8}
}
\altaffiltext{1}{Graduate School of Science and Engineering, Aoyama Gakuin University, 5-10-1 Fuchinobe, \\Sagamihara, Kanagawa 229-8558}
\email{yujin@phys.aoyama.ac.jp}
\altaffiltext{2}{The Institute of Physical and Chemical Research, 2-1 Hirosawa, Wako, Saitama 351-0198}
\altaffiltext{3}{Space Sciences Laboratory, University of California at Berkeley, 7 Gauss Way, Berkeley, \\California, 94720-7450, USA}
\altaffiltext{4}{LATT, Observatoire Midi-Pyr\'{e}n\'{e}es, UMR5572, CNRS-UPS, 14 av. E.Belin, 31400 Toulouse, France}
\altaffiltext{5}{Department of Physics, Tokyo Institute of Technology, 2-12-1 Ookayama, Meguro-ku, \\Tokyo, 152-8551}
\altaffiltext{6}{National Astronomical Observatory, 2-21-1 Osawa, Mitaka, Tokyo, 181-8588}
\altaffiltext{7}{Tsukuba Space Center, JAXA, 2-1-1 Sengen, Tsukuba, Ibaraki, 305-8505}
\altaffiltext{8}{Center for Space Research, MIT, 70 Vassar Street, Cambridge, Massachusetts, 02139, USA}
\altaffiltext{9}{INAF/IASF Bologna, Via Gobetti 101, 40129 Bologna, Italy}
\altaffiltext{10}{Noqsi Aerospace, LTd., 2822 South Nova Road, Pine, Colorado, 80470, USA}

%%% end:list of authors

%%% Please use the following style in case that sorting by 
%%% affilation is impossible. 
%
% \author{%
%   D-Firstname \textsc{D-Familyname}\altaffilmark{1}
%   E-Firstname \textsc{E-Familyname}\altaffilmark{1,2}
%   and
%   F-Firstname \textsc{F-Familyname}\altaffilmark{2}}
% \altaffiltext{1}{Address of Institute}
% \email{ddddd@xxx.xxx.xx.xx}
% \email{eeeee@xxx.xxx.xx.xx}
% \altaffiltext{2}{Address of Institute}

%% `\KeyWords{}' always has to be placed before `\maketitle'.
\KeyWords{stars: pulsars individual(SGR\,1806$-$20, SGR\,1900$+$14) --- X-rays: stars --- gamma rays: observations} %Do NOT move this preamble from here!

\maketitle

\begin{abstract}
We present the results of temporal and spectral studies of
the short burst (less than a few hundred milliseconds) from the soft gamma repeaters (SGRs)
1806$-$20 and 1900$+$14 using the HETE-2 samples.
In five years from 2001 to 2005,
HETE-2 detected 50 bursts which were localized to SGR\,1806$-$20 and
5 bursts which were localized to SGR\,1900$+$14.
Especially SGR\,1806$-$20 was active in 2004, and HETE-2 localized 33 bursts in that year.
The cumulative number-intensity distribution of SGR\,1806$-$20 in 2004 
is well described by a power law model with 
an index of $-1.1\pm0.6$. It is consistent
with previous studies but burst data taken in other years clearly give
a steeper distribution.
This may suggest that more energetic bursts could occur
more frequently in periods of greater activity.
A power law cumulative number-intensity distribution is also known
for earthquakes and solar flares.
It may imply analogous triggering mechanisms.
Although spectral evolution during bursts with a time scale of
$\gtrsim$ 20\,ms is not common in the HETE-2 sample, spectral softening due to the very rapid 
($\lesssim$ a few milliseconds) energy reinjection and cooling may not be excluded.
The spectra of all short bursts are well reproduced by a two blackbody function (2BB)
with temperatures $\sim 4$ and $\sim 11$\,keV.
From the timing analysis of the SGR\,1806$-$20 data, a time lag of
$2.2\pm0.4$\,ms is found between the 30-100\,keV and 2-10\,keV radiation bands.
This may imply (1) a very rapid spectral softening and energy
reinjection, (2) diffused (elongated) emission plasma along the
magnetic field lines in pseudo equilibrium with multi-temperatures, or
(3) a separate (located at $\lesssim 700$\,km) 
emission region of softer component (say, $\sim 4$\,keV) which could be
reprocessed X-rays by higher energy ($\gtrsim 11$\,keV)  photons from an emission region 
near the stellar surface.  
\end{abstract}

\section{Introduction}\label{introduction}
The soft gamma repeaters (SGRs) are sources of repeating short-duration ($\sim$100\,ms),
soft-spectrum (below $\sim$100\,keV) bursts.  Their activity periods are sporadic
and unpredictable, and are punctuated by long periods of inactivity.  More rarely,
the SGRs emit giant flares which last for minutes and display very hard X- and gamma-ray
spectra, extending into the MeV range.  The first burst 
from a soft gamma repeater (SGR) was detected
on January 7 1979 by experiments on the {\it Venera} 11, 
ICE, and Prognoz-7 spacecraft \citep{maz81,lar86}.  
This event was localized to a small region in the constellation Sagittarius, and was
accordingly named SGR\,1806$-$20.  In the following months, a second SGR (1900+14) was
discovered \citep{maz81}.
Around the same time, the first SGR giant flare occurred, on March 5 1979 \citep{maz79a, 
cli80, eva80, fen96} and was localized to
a supernova remnant in the Large Magellanic Cloud. That source is currently known as SGR\,0526$-$66.

For many years, little was known about the SGRs.
The giant flares were initially thought 
to be a special class of gamma-ray bursts (GRBs) \citep{maz81}, but since the SGRs
emitted small bursts repeatedly, they eventually came to be recognized as a separate phenomenon.  
In 1990's, quiescent soft X-ray observations ($\sim$ 1 -- 10\,keV) revealed that the
SGRs were quiescent, periodic soft X-ray emitters.  Their periods are 
of the order of seconds, and they display rapid spin-down rates \citep{kou98}.
Furthermore, since some of the SGRs appeared to be
associated with supernova remnants, they were regarded as a kind of X-ray pulsar.
(Today, these SNR-SGR associations have been questioned \citep{gae01},
and some indeed appear to be chance alignments, while other associations may be real.)

Ordinarily, isolated radio pulsars are thought to convert their rotational energy to radiation.
In contrast, the energy released in the persistent X-ray emission from SGRs is too
high to be explained in this way.
In addition, there seem to be no fluctuations in their rotational periods, nor 
any sign of a companion star, so that SGRs cannot be accreting 
neutron stars like the binary X-ray pulsars.
Rather, the SGRs are thought to be magnetars, i.e.
neutron stars with super-strong magnetic fields, \citep{dun92, pac92, tho95, tho96}.
Assuming that most of the spindown energy loss is due to
dipole radiation, the magnetic field strength should be $\sim 10^{15}$ G.

In the magnetar model, the surface of the neutron star
is heated and fractured by the movement and dissipation of the magnetic field.
The footpoints of the frozen-in magnetic field suddenly shift when a fracture occurs.
This generates Alfv\'{e}n waves which accelerate
electrons, which in turn radiate their energy in short bursts.
The giant flares are explained by global crustal fractures, which
lead to magnetic reconnection, and energization of the neutron
star magnetosphere, whose intense magnetic field contains the radiating electron-positron pair plasma
\citep{tho95, tho96}.
The other models propose that bursting activity
results from the heating of the corona by magnetic reconnection \citep{lyu03},
or the glitches in the p-stars \citep{cea2006}.

Today we know of four SGRs (1806$-$20, 1900$+$14, 0525$-$66, and 1627$-$41), as well
as three candidates: SGR\,1801$-$23 which emitted
two bursts detected by BATSE and the Interplanetary Network in 1997 \citep{cli00},
SGR\,1808$-$20 which emitted one burst detected by HETE-2 (the High Energy
Transient Explorer) in 2003 \citep{lam03}
and GRB/SGR\,050925 which was detected by {\it Swift} \citep{hol05}.
Compared with other bursting sources in our galaxy such as type I and type II
X-ray bursters, black hole X-ray transients, cataclysmic variables,
and novae, SGR bursts are the brightest.  Over the last $\sim$ 15 years,
SGR\,1806$-$20 and SGR\,1900$+$14 have been the most active, with $\sim$ 500 and $\sim$ 175 short
bursts observed, as well as one giant flare each.

Although the distances to SGR\,1806$-$20 and
SGR\,1900$+$14 have been inferred by various methods, they still 
remain very uncertain.
It was suggested earlier that the distance of SGR\,1806$-$20 is $d = 15.0$\,kpc
based on observations of molecular clouds along the line of sight \citep{cor97}.
A more recent study by \citet{cor04} proposed $d = 15.0_{-1.3}^{+1.8}$\,kpc based on
millimeter and infrared observations along G10.0$-$0.3.
Based on the measurements performed by observing the
radio afterglow of the giant flare on 27 December 2004,
two other estmates are published.
\citet{cam05} suggest that its distance is between 6.4\,kpc and
9.8\,kpc, while \citet{mcc05} simply infer a lower limit of 6\,kpc and
discuss that the most probable distance could be $d \sim 15$\,kpc.
If SGR\,1900$+$14 is associated with G42.8$+$0.6, the distance is
$\sim$ 5\,kpc \citep{hur99c}.
Alternatively, \citet{vrb00} found a compact cluster of massive stars
near SGR\,1900$+$14 and they suggested that the distance was between
12\,kpc and 15\,kpc if this was its birthplace.  In this paper, we assume distances
$d = 15$\,kpc for SGR\,1806$-$20 
and $d = 10$\,kpc for SGR\,1900$+$14.

The absorbing neutral hydrogen column densities along the lines of sight towards
SGR\,1806$-$20 and SGR\,1900$+$14 were investigated using ASCA data.
They turned out to be $\sim 6 \times 10^{22}$\,cm$^{-2}$ for SGR\,1806$-$20
\citep{son94} and $\sim 2 \times 10^{22}$\,cm$^{-2}$ for SGR\,1900$+$14
\citep{hur99c}.
The Galactic values in the directions of the sources are roughly $\sim$ 1.53 $\times$ $10^{22}$\,cm$^{-2}$
for SGR\,1806$-$20 which is marginally consistent with the ASCA result, 
and $\sim$ 1.57 $\times$ $10^{22}$\,cm$^{-2}$ for SGR\,1900$+$14 which is fully
consistent with the ASCA result \citep{dic90}.

In some studies, the spectra of the short SGR bursts have been well fitted by an optically thin thermal bremsstrahlung
(OTTB) spectrum (e.g., \cite{apt01}).
However, \citet{lar86} and \citet{fen94} reported that the spectra of bursts from
SGR\,1806$-$20 detected by
ICE were not well described by OTTB or other simple models.
\citet{oli04} reported that the spectra of an intermediate flare
(i.e. several seconds long) from SGR\,1900$+$14
detected by HETE-2 was not
described by OTTB at lower energies (below 15\,keV), but that
they were best described by a two blackbody function (2BB).
Similar results were reported for the short bursts from SGR\,1900$+$14 \citep{fer04}
and SGR\,1806$-$20 \citep{nak05}.
\citet{got06} reported that the spectrum of the large outburst from SGR\,1806$-$20
was well fitted by OTTB, and they also noted the possibility of 2BB.
\citet{fer04} found that fits to the short bursts from SGR\,1900$+$14 
by a power law with an exponential cutoff gave acceptable results.

Some authors reported that there was weak or no spectral evolution
for SGR bursts \citep{kou87, fen94}.
In other observations, however, spectral softening during
short bursts was reported from SGR\,1806$-$20 \citep{str98},
in a $\sim$3.5\,s long burst from SGR\,1900$+$14 \citep{ibr01},
and in a $\sim$9\,s long burst from SGR\,0526$-$66 \citep{gol87}.
Furthermore, significant spectral evolution was reported for the weaker bursts
from SGR\,1806$-$20 \citep{got04, got06}.

Here, we report the results of the temporal and spectral analyses of
the short bursts from SGR\,1806$-$20 and SGR\,1900$+$14,
whose durations are less than a few hundreds of milliseconds,
using the data of the Wide-Field X-ray Monitor
(WXM; 2-25\,keV energy band; \cite{shi03})
and French Gamma Telescope
(FREGATE; 6-400\,keV energy band; \cite{att03}) instruments on-board HETE-2.
We discuss the spectral evolution, the emitting region, the cumulative
number-intensity distribution and the spectral modeling of the short bursts,
and confront our results with theoretical models.

\section{Observations}\label{observation}
The primary purpose of HETE-2 is to investigate cosmic high-energy
transient phenomena such as GRBs \citep{ric03}.
The scientific instruments on-board HETE-2 always point in the
anti-solar direction and
the Galactic center region comes into the field of view
during the summer season; at that time
bursts from SGR\,1806$-$20 and SGR\,1900$+$14 are best observed.

Most of the bursts from the SGRs trigger the instruments in the 
6-80\,keV or 30-400\,keV energy range, on a 20\,ms time scale,
while some bursts trigger them in the 2-25\,keV energy range on a 320\,ms time scale.
Based on these criteria, HETE-2 triggered on 181 events
in the summer periods from June 18, 2001 through August 7, 2005.
We localized these events 
using data from the WXM, and found that 62 were from SGR\,1806$-$20 and 6 were
from SGR\,1900$+$14, while 113 events were out of the WXM field of
view ($> 35$ degrees off axis) or too weak to be localized ($< 10^{-7}$\,ergs\,cm$^{-2}$).
Table \ref{evt_sum} contains more details.
Although these nonlocalized events
were presumably bursts from SGR\,1806$-$20 or SGR\,1900$+$14   
considering the field-of-view of the instruments 
and their short durations ($\lesssim$ a few $\times$100\,ms)
we do not include them in the current analyses.

In this paper, we call {\it events} the set of data recorded after triggers by 
the instruments on the spacecraft.
Some events contain a single burst, and others are better described as containing
several separate bursts (see below).
The localized events include two intermediate flares
(HETE-2 trigger numbers  \#1576 and \#3800),
which we exclude from our analysis.  
We also exclude the following events:
HETE-2 trigger \#1560, a burst from SGR\,1806$-$20,
in which the time-tagged photon data from the WXM were lost,
three events in which the FREGATE time-tagged photon data were lost
(\#3290, \#3801 and \#3845), 
five events for which the downlinked tagged times were out of order (\#2768, \#2810, \#3282, \#3404 and \#3418), 
and three events whose data were corrupted due to strong 
X-ray sources like {\it Sco-X1} in the field-of-view, which made 
detailed analysis very difficult 
(\#3273, \#3274 and \#3349).
Therefore we retain
49 events containing 50 bursts from SGR\,1806$-$20,
and 5 events containing 5 bursts from SGR\,1900$+$14.

\begin{figure}
 \begin{center}
  \FigureFile(160mm,210mm){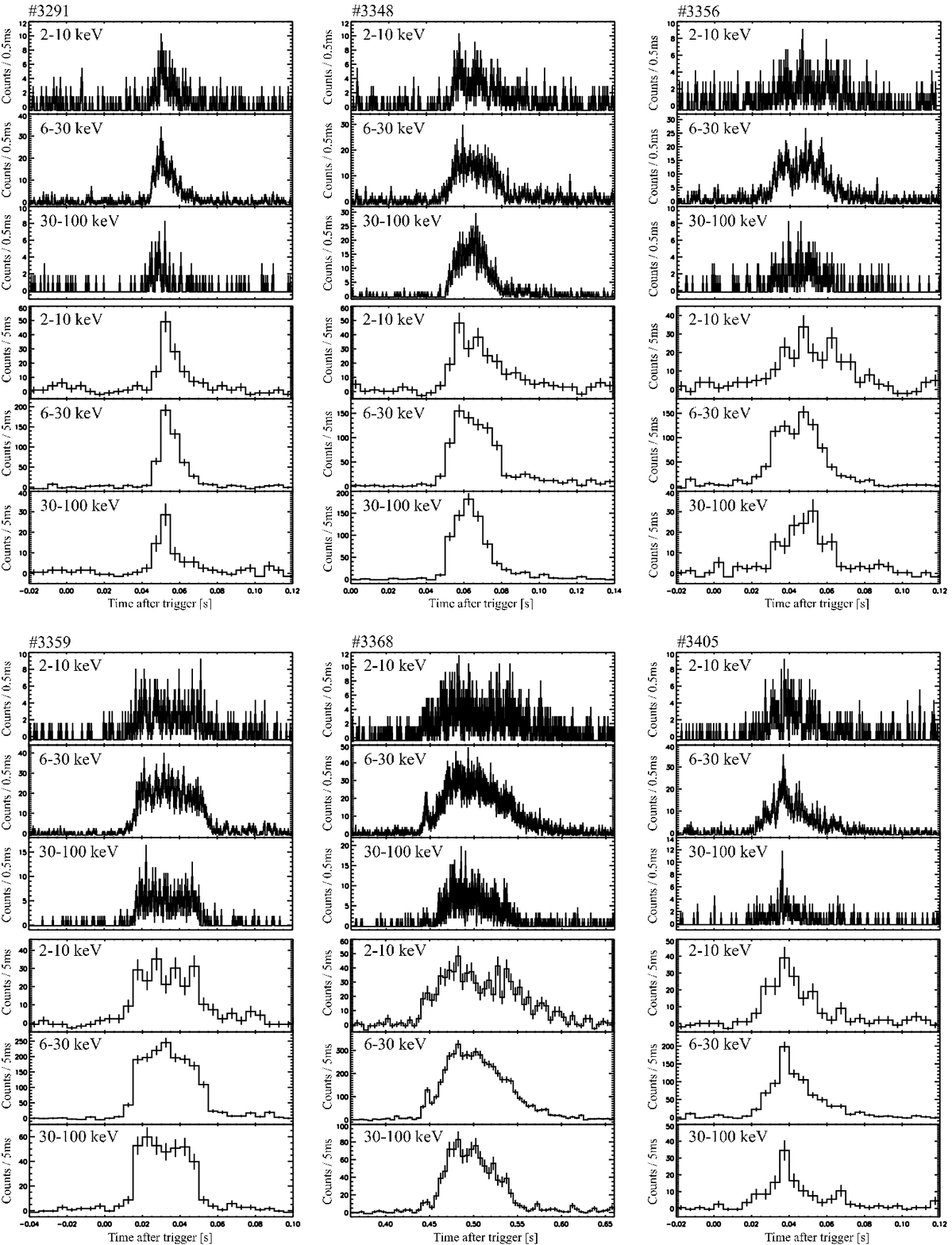}
 \end{center}
 \caption{Examples of the light curves binned with 0.5\,ms and 5\,ms resolution
in various energy bands for the bright single peaked bursts.}\label{bright_single_flares}
\end{figure}

\begin{figure}
 \begin{center}
  \FigureFile(160mm,124mm){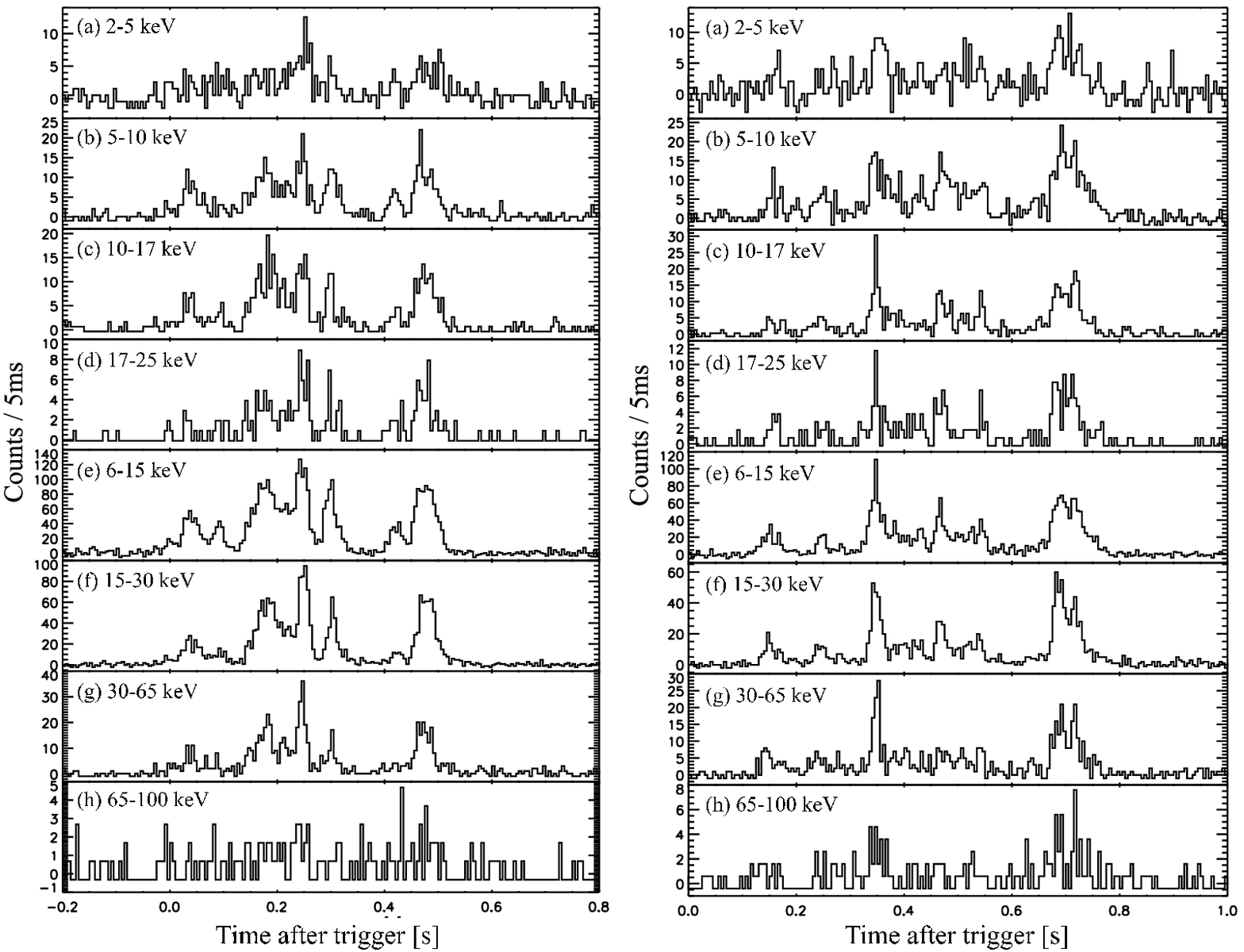}
 \end{center}
 \caption{Time history of \#3259 ({\it left}) and \#3303 ({\it right}) observed
 by the WXM in the energy bands 2-5\,keV (a), 5-10\,keV (b), 10-17\,keV (c)
 and 17-25\,keV (d), and by FREGATE in the 6-15\,keV (e),
 15-30\,keV (f), 30-65\,keV (g) and 65-100\,keV (h) energy bands with 5\,ms time bins.}\label{lc_3259_3303}
\end{figure}

To compare HETE-2 results with previous studies,
event classification is done using methods similar to those in \citet{gog01}.
The events are classified into  
single peaked bursts and multiple peaked  bursts 
using light curves with 5\,ms binning.
Examples are given in figure \ref{bright_single_flares}
for the single peaked bursts and in figure \ref{lc_3259_3303} for
the multiple peaked bursts.
22 out of 50 bursts from SGR\,1806$-$20 are single peaked and 28 are multiple peaked.
From SGR\,1900$+$14, four single peaked and one multiple peaked bursts were detected.
For the multiple peaked bursts, if the interval between the first peak
and the second peak is greater than $T_{\mathrm{90}}$ for the first peak
($T_{\mathrm{90}}$ is defined as the time to accumulate between
5 \% and 95 \% of the observed photons),
the second peak is treated as another burst.
If we find two bursts within one event, we distinguish
those bursts using the letters A and B.
There are five triggered events which each contain two bursts from SGR\,1806$-$20
(\#2800, \#3347, \#3365, \#3368 and \#3399);
only one of them (\#2800) gives 
a significant localization to  SGR\,1806$-$20 for each burst,
and is included in the sample for the analyses. 
The number of localized events and bursts employed in our
analyses  are summarized in table \ref{evt_sum}.
Note that SGR\,1806$-$20 became very active in 2004 and 40 
events triggered  in the summer of 2004.
In contrast, SGR\,1900$+$14 has been in a relatively quiescent phase since HETE-2 launch.
All the analyzed bursts are listed in tables \ref{flare_list1} and \ref{flare_list2},
in which the single peaked bursts are indicated by $s$ and the multiple peaked ones by $m$.

\begin{figure}
 \begin{center}
  \FigureFile(80mm,83mm){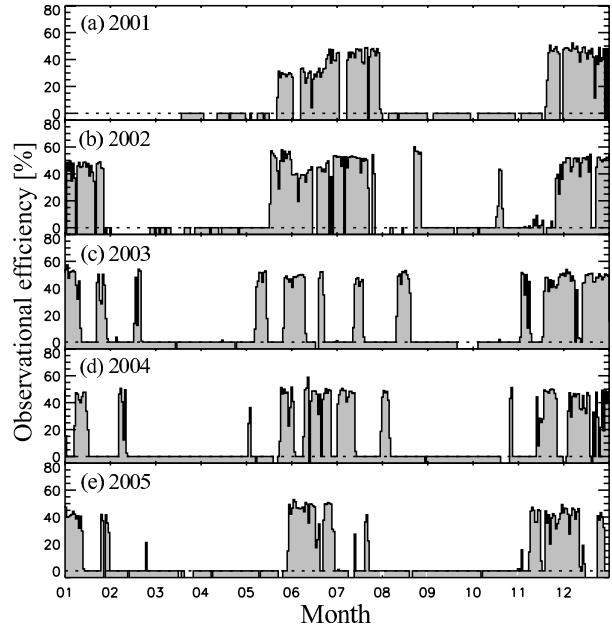}
 \end{center}
 \caption{Time history of the observational efficiency for each day (UT)
 of the WXM for SGR\,1806$-$20.
 The blank regions without hatching represent periods
 in which aspect or housekeeping data were lost (e.g. because of a downlink problem).}\label{amount_eff1}
\end{figure}

\begin{figure}
 \begin{center}
  \FigureFile(80mm,83mm){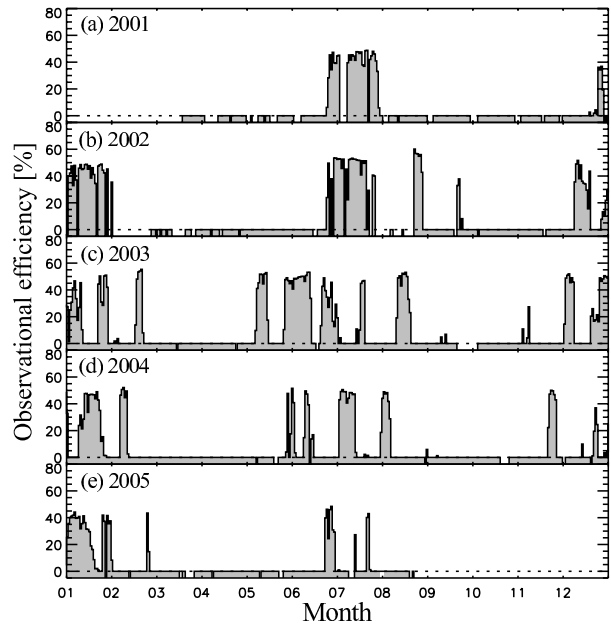}
 \end{center}
 \caption{Same plot as figure \ref{amount_eff1} for  SGR\,1900$+$14.}\label{amount_eff2}
\end{figure}

The observational efficiencies were estimated for each source
from June 2001 to August 2005.
First, we calculated the fraction of time when the high voltage was on ($\varepsilon_{\mathrm{hv}}$),
since HETE-2 performs observations only during orbit night. 
Then we derived the fraction of the time when each SGR was in the
field-of-view of the 35 degree WXM ($\varepsilon_{\mathrm{fov}}$).
The observational efficiency is given by 
$\varepsilon_{\mathrm{hv}} \times \varepsilon_{\mathrm{fov}}$
for each source.
The WXM on-board HETE-2 is operated so that the high voltage is turned off during
the periods when the earth is in the field-of-view.
Therefore $\varepsilon_{\mathrm{hv}}$ take into account the earth occultations.
Figure \ref{amount_eff1} and figure \ref{amount_eff2} show time histories of the
observational efficiencies with one day resolution for SGR\,1806$-$20 and SGR\,1900$+$14, respectively.
The blank regions without hatching represent periods 
in which aspect or housekeeping data were lost (e.g. because of a downlink problem).
With these observational efficiencies, the expected number of bursts 
was estimated for each year.
Table \ref{corrected_flares} shows these numbers, as well as the observing time in kiloseconds.

\section{Analysis}
\subsection{Temporal Analysis}
\subsubsection{Distribution of $T_{\mathrm{90}}$}
To study the burst durations, we calculated $T_{\mathrm{90}}$ for all the
samples from SGR\,1806$-$20 and SGR\,1900$+$14.
Our samples do not have too bright burst for which the dead time 
needs to be considered for analyses.

\begin{figure}
 \begin{center}
  \FigureFile(160mm,81mm){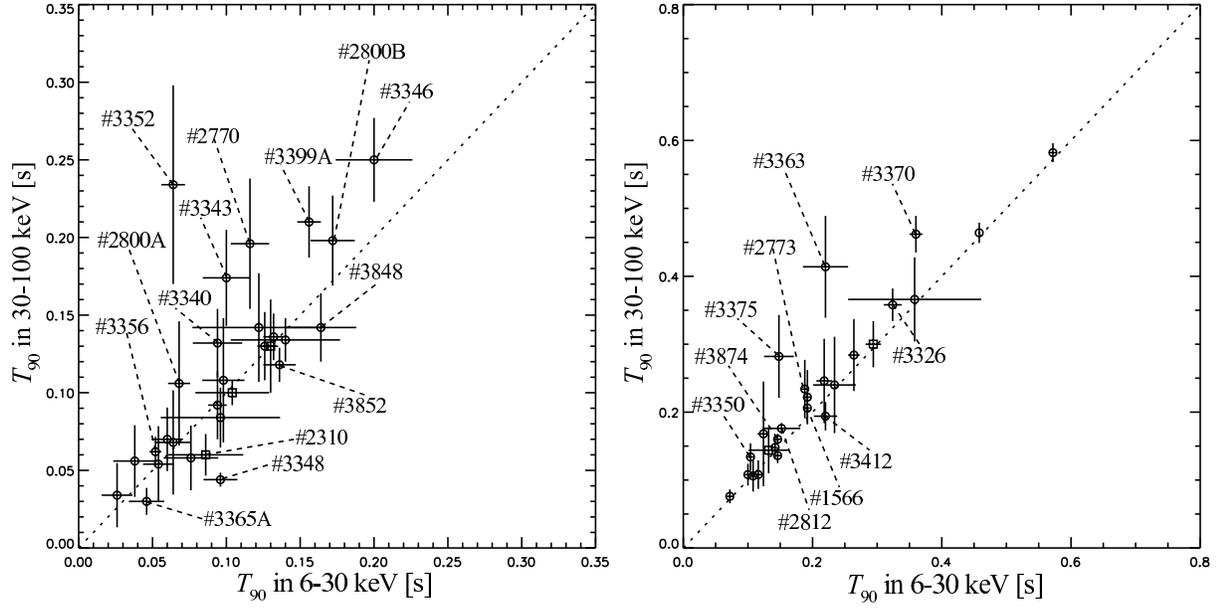}
 \end{center}
 \caption{Distribution of $T_{\mathrm{90}}$ durations for the single peaked
 bursts (left) and multiple peaked bursts (right). The circle and square
 show the $T_{\mathrm{90}}$ durations for SGR\,1806$-$20 and SGR\,1900$+$14,
 respectively. The dotted lines indicate equal
 6-30\,keV and 30-100\,keV $T_{\mathrm{90}}$. The quoted errors are for 90 \% confidence.}\label{t90_dist}
\end{figure}

The $T_{\mathrm{90}}$ durations are shown in tables \ref{flare_list1} and \ref{flare_list2}.
Figure \ref{t90_dist} shows the relation between
the $T_{\mathrm{90}}$ durations in the 6-30\,keV and 
30-100\,keV energy ranges for the single peaked bursts (left) and the multiple
peaked bursts (right);  the dotted lines indicate equal $T_{\mathrm{90}}$ durations in both energy ranges.
Outlying points indicate that the burst might have spectral evolution.
The circles and squares show the $T_{\mathrm{90}}$ durations for SGR\,1806$-$20
and SGR\,1900$+$14, respectively.
Although most of the bursts lie on the dotted lines
(i.e. they appear to be consistent with no spectral evolution),
we find that 13 single peaked bursts from SGR\,1806$-$20
(indicated by $b$ in table \ref{flare_list1})
and one single peaked burst from SGR\,1900$+$14
(indicated by $b$ in table \ref{flare_list2})
lie far from them.
We also find that 10 multiple peaked bursts from SGR\,1806$-$20
(indicated by $c$ in table \ref{flare_list1})
are significantly inconsistent with the dotted line.

We have also studied the hardness ratios for the bursts from
SGR\,1806$-$20 and SGR\,1900$+$14 as a function of time.
The hardness ratio {\it HR} is defined by {\it HR} $ = (H-S)/(H+S)$
where $S$ is the 6-30\,keV count rate and $H$ is the 30-100\,keV count rate.
The hardness ratios are summarized in figure \ref{hr_plots1}.
We find that three short bursts
(\#3340, \#3365A and \#3874 which are indicated by $d$ in table \ref{flare_list1})
have clear spectral softening, while another three
(\#2800A, \#3356 and \#3370 which are indicated by $e$ in table \ref{flare_list1})
might have a hard component at end of the burst.
Although one short burst \#3348 seems to show a marginal spectral softening,
the last three bins of its hardness ratios have large uncertainties.
Then it is omitted from the above group.

\begin{figure}
  \begin{center}
   \FigureFile(160mm,85mm){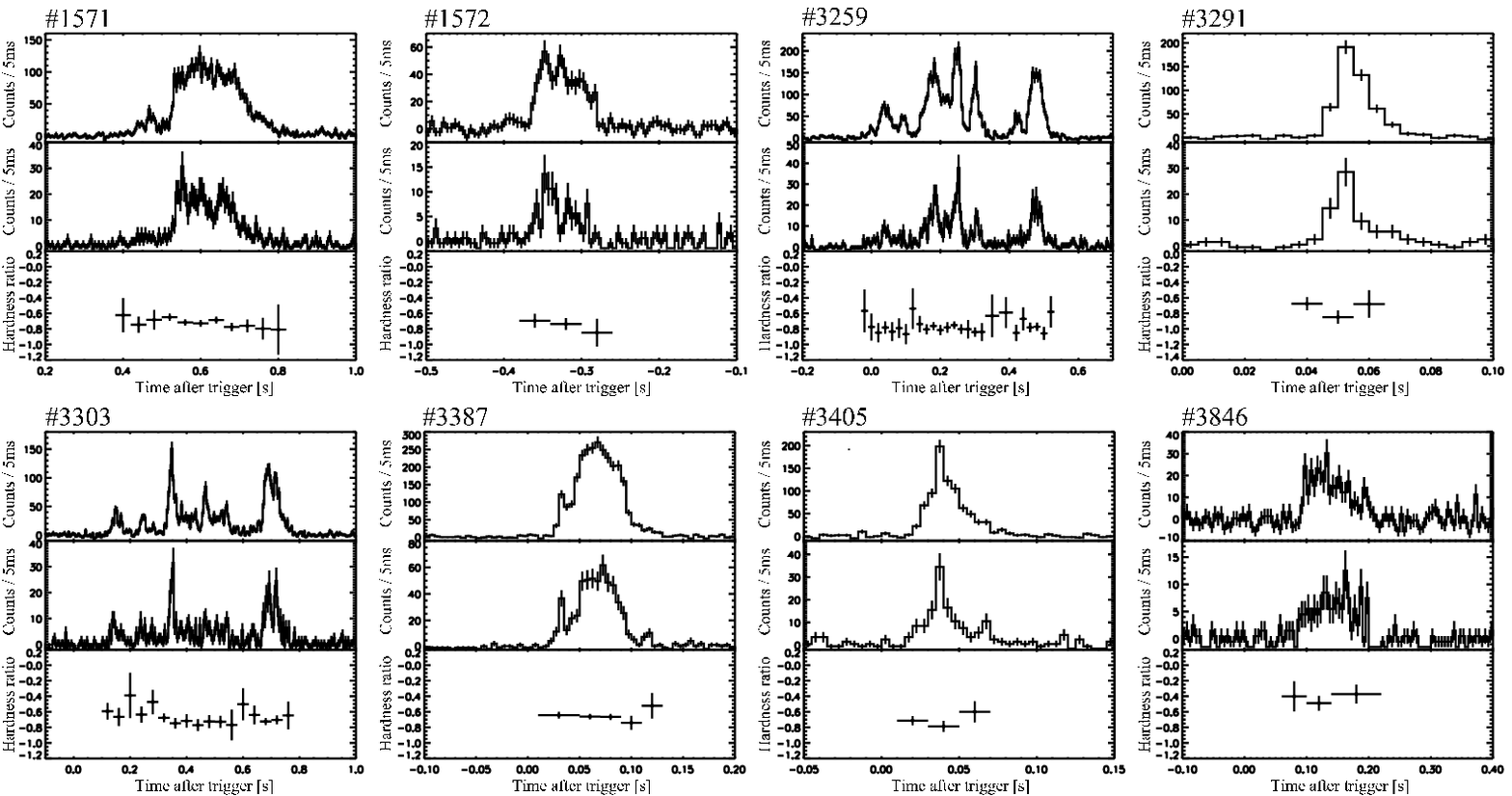}
  \end{center}
  \caption{Light curves observed by {\it FREGATE} in the 6-30\,keV 
 ({\it top panels})  and the 30-100\,keV ({\it middle panels}) energy bands,
 and the time resolved hardness ratio ({\it bottom panels});
 these are for the short bursts which lie on the dotted lines described
 in figure \ref{t90_dist} (i.e. no spectral evolution).
 The quoted errors are for 68 \% confidence.}\label{hr_plots3}
\end{figure}

As a control, we also calculated the hardness ratios
for eight short bursts which clearly lie on the dotted lines
(indicated by $a$ in table \ref{flare_list1} and table \ref{flare_list2}).
These hardness ratios are shown in figure \ref{hr_plots3} and no spectral evolution
is seen for them.

\subsubsection{Time Lag}\label{time_lag}
Some bursts seem to display delayed soft emission (e.g. \#3259 - see figure \ref{lc_3259_3303}).
To investigate systematically if a delay appears in the softer energy band,  
the time lag $T_{\mathrm{lag}}$ between the 2-10\,keV and 30-100\,keV time histories, and between
the 6-30\,keV and 30-100\,keV time histories, were evaluated by cross-correlating them.
In our definition, the time lag is positive when the soft emission delay
from the hard emission.
Here the boundary energy of 30\,keV was chosen because it is approximately 
the dividing point between the spectral components in a two blackbody function
(see $\S$ \ref{spec_two_bb} and $\S$ \ref{spectral_properties}).

In order to select statistically significant bursts for this analysis,
we proceeded in the following way.
First, background-subtracted light curves were generated with 5\,ms time bins to select 
bursts using the peak count rates in the 2-10\,keV, 6-30\,keV, and 30-100\,keV energy ranges.
To estimate a correct peak count rate, the phasing of the time bins
(i.e. the beginning time of $t = 0 $ bin) is important.
We determined the phase by shifting the beginning time of $t = 0$ bin
with 0.25 time interval so that the peak count rate in the light curve
got the maximum value.
Using this method, the 5\,ms light curves and peak count rates were derived.
The cross-correlation coefficients were calculated for each burst.
We searched the requirements of peak count rate by investigating
whether the peak of the cross-correlation coefficients was clearly seen or not
for each burst.
As a result, we required the peak count rates to be $>$10$\sigma_{\mathrm{1}}$,
$>$20$\sigma_{\mathrm{2}}$ and $>$22$\sigma_{\mathrm{3}}$ in the 2-10\,keV,
6-30\,keV, and 30-100\,keV energy ranges, respectively, where
$\sigma_{\mathrm{1}} = 0.4$, $\sigma_{\mathrm{2}} = 0.6$ 
and $\sigma_{\mathrm{3}} = 0.2$\,counts\,ms$^{-1}$ are the standard deviations of
the background in each band.
We did not employ the bursts showing clear spectral 
softening (\#3340, \#3365A and \#3874)
and omitted the burst with a marginal spectral softening (\#3348) either 
to avoid affecting the time lag analysis.

After these selections, only 16 bursts from SGR\,1806$-$20 survived for
the $T_{\mathrm{lag}}$ analysis, which are indicated by $l$ in table \ref{flare_list1}.
There were no bursts satisfying the above criteria for SGR\,1900$+$14.

To calculate the cross-correlation coefficients, background-subtracted
light curves with 0.2\,ms time bins were generated 
for each burst (their phases were determined with 0.01\,ms time interval).
We then smoothed the light curves to reduce fluctuations.
The moving average $N(t)'$ at a given time was calculated by
$(N(t)+N(t-{\Delta}t))/2$, where $t$ is the time, ${\Delta}t$ is the time resolution
of the light curve and $N(t)$ and $N(t-{\Delta}t)$ are the count rates.
Even after this smoothing, however,
the cross-correlation coefficients still displayed large fluctuations.
We therefore made an ensemble average by assuming that all the bursts were samples from
the same parent population; in other words, the same physical process controls
the radiation for all short bursts.

\begin{figure}
 \begin{center}
  \FigureFile(80mm,80mm){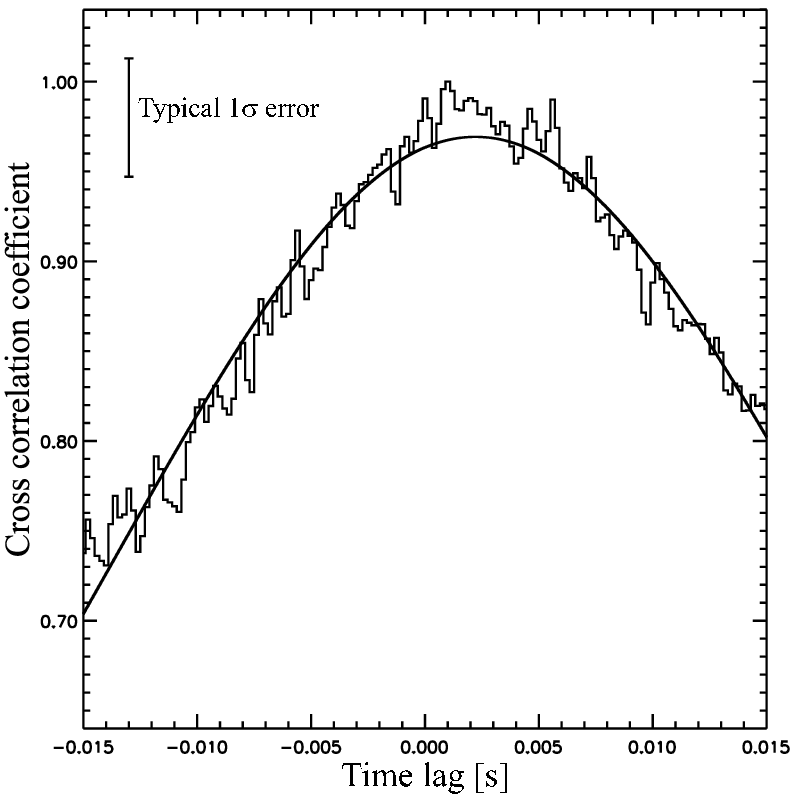}
 \end{center}
 \caption{Cross correlation coefficients between the 2-10\,keV and 30-100\,keV
 time histories.}\label{cc_summary}
\end{figure}

The resultant cross-correlation coefficients were fitted with
a gaussian model, a symmetric exponential component model, and a lorentzian model,
for all of which a time lag was derived to be positive.
Here we employ the lorentzian model because it best 
reproduces the shape of the cross correlation coefficient.
The best-fit $T_{\mathrm{lag}}$ is $2.2\pm0.4$\,ms between the 2-10\,keV
and 30-100\,keV time histories, and $1.2\pm0.3$\,ms between 6-30\,keV and 30-100\,keV.
Here the quoted errors are total uncertainties 
($\sqrt{{\sigma_{\mathrm{stat}}}^2 + {\sigma_{\mathrm{sys}}}^2}$), 
considering time resolutions of the light curves (0.2\,ms) and the instruments
(256\,$\mu$s for the WXM \citep{shi03} and 6.4\,$\mu$s for the FREGATE
\citep{att03}) and therefore systematic uncertainty ($\sigma_{\mathrm{sys}}$) of 
0.3\,ms and 0.2\,ms 
for the time lag between the 2-10\,keV
and 30-100\,keV bands and between 6-30\,keV and 30-100\,keV bands respectively.
Figure \ref{cc_summary} shows the cross correlation coefficients between
2-10\,keV and 30-100\,keV, 
and the typical 1$\sigma$ error is presented in the figure.

For each of the 16 bursts, 100 light curves are simulated 
for both the 2-10\,keV and 30-100\,keV band 
to investigate if the derived lag is just due to fluctuations of counts. 
Applying the same method for observational data,
time lags are found to be distributed above 0.6\,ms in these simulated data.
Therefore we conclude that the nonzero time lag is real.

\begin{figure}
 \begin{center}
  \FigureFile(80mm,80mm){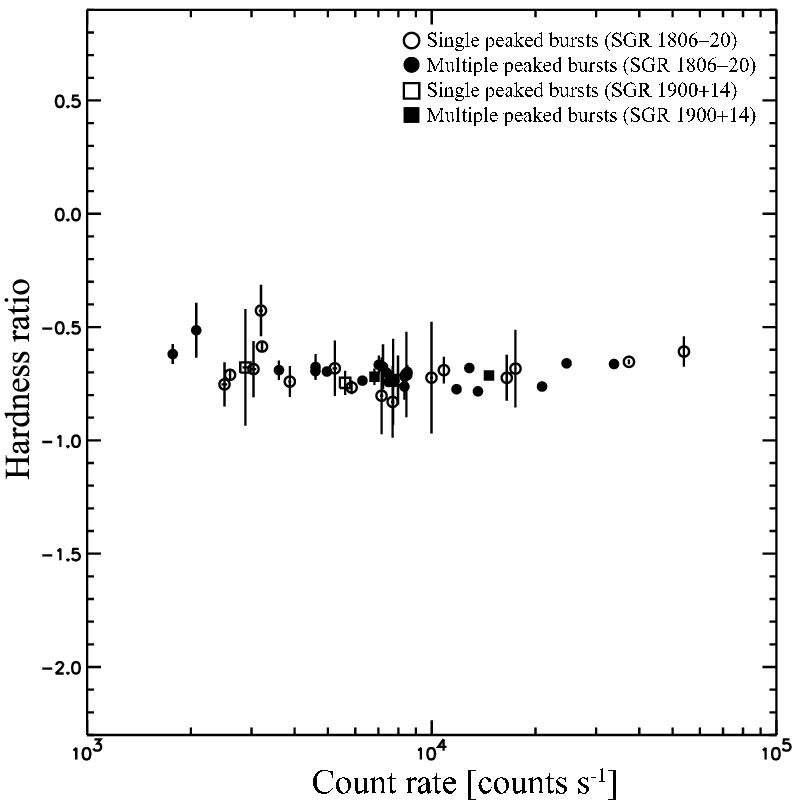}
 \end{center}
 \caption{Relation between the 6-100\,keV count rate and the hardness ratio.
 11 bursts observed with the different gain configuration (indicated by $h$ in
 table \ref{flare_list1}) are not employed in this figure.}\label{hardness_intensity}
\end{figure}

Using the peak count rates derived above, the relation between the
6-100\,keV fluence and the hardness ratio was also 
investigated (figure \ref{hardness_intensity}).
The hardness ratio {\it HR} was defined as {\it HR} $ = (H - S)/(H + S)$ where $S$
is the 6-30\,keV count rate and $H$ is the 30-100\,keV count rate.
11 bursts observed with the different gain configuration (indicated by $h$ in
table \ref{flare_list1}) are not employed in this figure.
There seems to be no significant correlation.
This result is consistent with previous studies \citep{fen94, got04, got06}.

\subsection{Spectral Analysis}\label{spectral_analyses}
The following eight functions were used as models 
for spectral fitting of all the samples; 
1) a power law model (PL), 
2) a power law with an exponential cutoff (PLE), 
3) a single blackbody (BB),
4) a two blackbody function (2BB),
5) a disk-blackbody model (disk-BB), 
6) an optically thin thermal bremsstrahlung model (OTTB),
7) OTTB with BB model (BB$+$OTTB), and 
8) PL with BB model (BB$+$PL). 

If we add a photoelectric absorption to the 2BB model,
we find that the column densities are consistent with zero (i.e., no absorption).
On the other hand, if we adopt the photoelectric absorptions derived from
the ASCA observations of the quiescent emission (see $\S$ \ref{introduction}),
we find unacceptable results for a few cases.
Considering these results, we perform the spectral analyses 
using 2BB with a photoelectric absorption model in which the Galactic
absorptions are adopted as the lower limit values.

The single component models (PL, BB and OTTB) are rejected,
while BB$+$OTTB, PLE, disk-BB and BB$+$PL give acceptable fits with a few exceptions, 
and 2BB gives even better fits for all cases.

Because of the effects by problems of intercalibration (the reason is under investigation),
the FREGATE spectra below 20\,keV are somewhat different from those of the WXM.
For this reason, the FREGATE spectra below 20\,keV are not employed
in the joint fitting.
From the temporal analysis, we know that almost no emission is seen 
above 100\,keV in most cases; 
thus we employ the 2-25\,keV energy range for the WXM, and 20-100\,keV for FREGATE.

We define the foreground time region as a burst period, and the background
time region as a steady emission level (i.e. before and after the burst).
The foreground time regions which included the whole burst were determined by eye.
The background was selected from regions 8 seconds long 
at least both before and after the burst (foreground region).
For two bursts, \#3259 and \#3303, time-resolved spectral analysis was done,
because they have good enough statistics (greater than 3000\,counts, 6-30\,keV), and
longer durations ($\gtrsim300$\,ms) than those of typical bursts.
We used the 2BB model for these time-resolved fits.
Their time histories are shown in figure \ref{lc_3259_3303} ($t=0$ corresponds to the trigger time)
and the best-fit spectral parameters are summarized in 
table \ref{spc_result_3259} and table \ref{spc_result_3303}.

\subsubsection{BB$+$OTTB with Absorption}
In some earlier studies, the hard X-ray spectra of SGR bursts were found to be
well described by an OTTB function.
Therefore we tried OTTB fits, but found that they gave poor results in many cases.
The discrepancies were prominent in the lower energy region for the  WXM data.
Thus we introduced a BB component to describe the lower energies. 
This model gives better fits, but it requires a large absorption 
column to reproduce the steep turnover in the lower energy band.
We find that all bursts are well reproduced by BB$+$OTTB except for \#3405.
15 out of 55 bursts do not require a BB component.
For the remaining 40 bursts, if we fit the spectra only with OTTB,
we have an excess below 5\,keV.
The photoelectric absorption $N_{\mathrm{H}}$ is $\sim 10^{23}$\,cm$^{-2}$,
which is greater than the Galactic value.
The best-fit spectral parameters are summarized in table \ref{spc_list_bb_ottb_1} and
table \ref{spc_list_bb_ottb_2}.
The quoted errors represent the 68 \% confidence level for fluxes and the 90 \%
confidence level for other parameters.

The BB temperatures $kT_{\mathrm{BB}}$ are distributed around $\sim 5$\,keV and
the emission radii $R_{\mathrm{BB}}$ are distributed around $\sim 23$\,km
(at 15\,kpc for SGR\,1806$-$20 and at 10\,kpc for SGR\,1900$+$14).
Alternatively $R_{\mathrm{BB}}' = R_{\mathrm{BB}}(d'/\{15\,{\mathrm{kpc}}, 10\,{\mathrm{kpc}}\})$
at an arbitrary distance $d'$.
The OTTB temperatures $kT_{\mathrm{OTTB}}$ are distributed around $\sim$ 27\,keV
in most cases, which is consistent with previous studies (e.g., \cite{apt01}).
However \#3399A has a significantly smaller temperature, $kT_{\mathrm{OTTB}} \sim 2$\,keV.

\subsubsection{PLE with Absorption}
Because the OTTB model is unacceptable in many cases, we tried the PLE model;
PLE can be regarded as an extension of the simple OTTB.
The spectral parameters are summarized in table \ref{spc_list_cpl_1} and
table \ref{spc_list_cpl_2}.
The quoted errors are again 68 \% confidence level for fluxes and 90 \%
confidence level for other parameters.
We find that all bursts except for \#1571 and \#3405 are well reproduced by PLE.
For \#1571 and \#3405, the values of the reduced chi-squared
are rather large, but still acceptable.
The photoelectric absorption $N_{\mathrm{H}}$ is $\sim 10^{23}$\,cm$^{-2}$, again
greater than the Galactic value.
The spectral indices $\alpha$ are distributed around $\sim$ 0.5, which
is relatively hard compared to GRB indices.
The cutoff energies $E_{\mathrm{0}}$ are distributed around $\sim$ 13\,keV.
These spectral parameters are consistent with previous observations \citep{fer04}.

\subsubsection{2BB}\label{spec_two_bb}
Some studies report that the spectra of certain SGR bursts are well
reproduced by the 2BB model.
In a previous study of HETE-2 samples, this model gave good fits \citep{nak05}.
In the current study, which uses all the events available to date, 2BB
again describes the observed spectra well.
The spectral parameters are summarized in table \ref{spc_list_2bb_1} and
table \ref{spc_list_2bb_2}.
The quoted errors are 68 \% confidence for fluxes and 90 \% confidence for other parameters, as before.
Using 2BB, we get acceptable fits in all cases.
2 out of 55 bursts do not require a higher blackbody component.
Only the upper limit of the photoelectric absorption $N_{\mathrm{H}}$ is determined
for most cases (i.e. which is consistent with zero).
The lower blackbody temperatures $kT_{\mathrm{LT}}$ and their radii $R_{\mathrm{LT}}$
are distributed around $\sim$ 4.2\,keV and $\sim$ 27\,km
(at 15\,kpc for SGR\,1806$-$20 and at 10\,kpc for SGR\,1900$+$14), respectively.
The higher blackbody temperatures $kT_{\mathrm{HT}}$ are generally distributed around
$\sim$ 11\,keV, but  for
\#1572, \#2310, \#3352, \#3355, \#3378 and \#3851, we can only obtain lower limits.
The radii of the higher blackbody temperatures $R_{\mathrm{HT}}$ are distributed
around $\sim$ 4.5\,km (at 15\,kpc for SGR\,1806$-$20 and at 10\,kpc for SGR\,1900$+$14).
Again $R_{\{{\mathrm{LT, HT}}\}}' = R_{\{{\mathrm{LT, HT}}\}}(d'/\{15\,{\mathrm{kpc}}, 10\,{\mathrm{kpc}}\})$
at an arbitrary distance $d'$.

\subsubsection{disk-BB with Absorption}
We tried the disk-BB function because it can be regarded as an extended case of 2BB.
The spectral parameters are summarized in table \ref{spc_list_diskbb_1} and
table \ref{spc_list_diskbb_2}, where the quoted errors are 68 \% confidence for fluxes and 90 \%
confidence for other parameters.
We find that almost all bursts are well reproduced by disk-BB with some exceptions.
For \#3303, we find a deficit of photons around $\sim$ 30\,keV.
\#1571 and \#3368B show an excess below $\sim$ 5\,keV, while \#3326 shows an excess above $\sim$ 50\,keV.
The photoelectric absorption $N_{\mathrm{H}}$ is $\sim 10^{23}$\,cm$^{-2}$,
which is greater than the Galactic value.
The temperatures at the inner disk radius $T_{\mathrm{in}}$ are distributed
around $\sim$ 10\,keV, and the values of $R_{\mathrm{in}}\sqrt{\cos\theta}$ are
distributed around $\sim$ 4.5\,km (at 15\,kpc for SGR\,1806$-$20 and at 10\,kpc for SGR\,1900$+$14)
where $R_{\mathrm{in}}$ denotes the inner disk radius and $\theta$ denotes the angle of the disk.
As before, $R_{\mathrm{in}}'\sqrt{\cos\theta} = R_{\mathrm{in}}\sqrt{\cos\theta}(d'/\{15\,{\mathrm{kpc}}, 10\,{\mathrm{kpc}}\})$
at an arbitrary distance $d'$.

\subsubsection{BB$+$PL with Absorption}
The 2BB or disk-BB fits suggest that reprocessed X-ray emission
may be present in the spectra.
Another possibility is inverse Compton scattering by a medium such as a hot plasma.
The spectral parameters are summarized in table \ref{spc_list_bb_pow_1} and
table \ref{spc_list_bb_pow_2}.
The quoted errors are 68 \% confidence for fluxes and 90 \% confidence for other parameters.
We find that all bursts  are well reproduced by BB$+$PL, except for \#3326
which has an excess below 5\,keV.
The photoelectric absorption $N_{\mathrm{H}}$ is order of $\sim 10^{23}$\,cm$^{-2}$,
which is greater than the Galactic value.
The blackbody temperatures $kT_{\mathrm{BB}}$ are distributed
around $\sim$ 5.5\,keV and the blackbody radii $R_{\mathrm{BB}}$ are distributed
around $\sim$ 13\,km (at 15\,kpc for SGR\,1806$-$20 and at 10\,kpc for SGR\,1900$+$14).
$R_{\mathrm{BB}}' = R_{\mathrm{BB}}(d'/\{15\,{\mathrm{kpc}}, 10\,{\mathrm{kpc}}\})$
at an arbitrary distance $d'$.
The spectral indices $\alpha$ for the PL component are distributed around $\sim$ 2.1.

\section{Discussion}
\subsection{Cumulative Number-Intensity Distribution for SGR\,1806$-$20}\label{cumulative}
The cumulative number-intensity distribution for 2-100\,keV fluences was calculated
for the short bursts from SGR\,1806$-$20.
The fluences were derived from the 2BB fits.
Poor statistics of the single peaked bursts did not allow us to
investigate their distributions separately from those of the multiple peaked bursts.
Therefore we investigate the distribution using both burst samples.

\begin{figure}
 \begin{center}
  \FigureFile(80mm,81mm){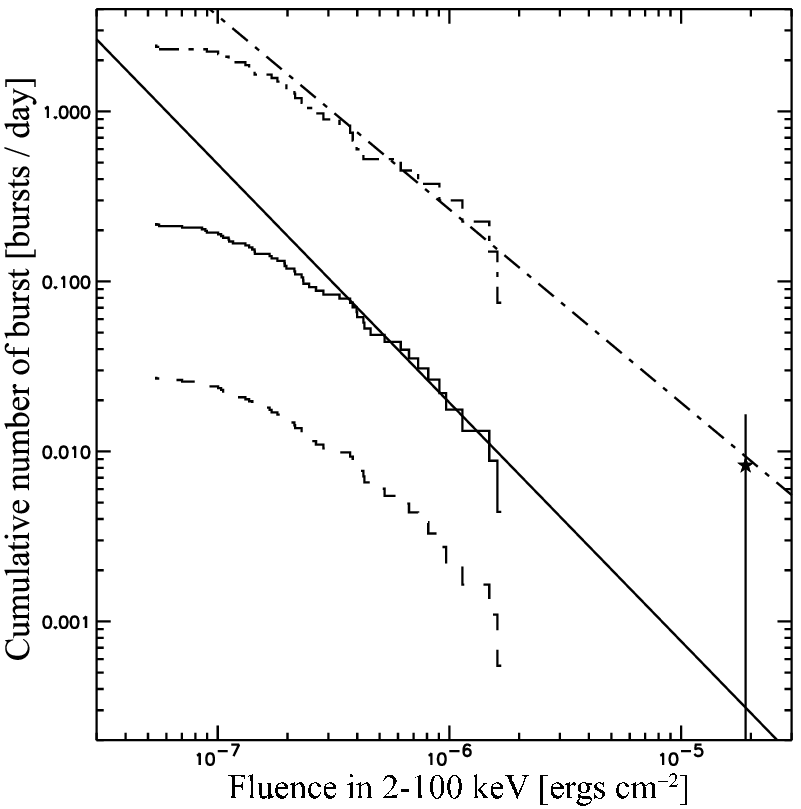}
 \end{center}
 \caption{Cumulative 2-100\,keV number-intensity distribution for SGR\,1806$-$20.
 The dashed line represents the observational data.
 The solid stepwise line represents the data corrected for
 observational efficiency, and the solid straight line represents
 the fit to it.
 The dot-dashed stepwise line shows the corrected distribution using
 the data in 2004 and the dot-dashed straight line shows its fitting result.
 The star symbol is for the intermediate flare.}\label{number_intensity_fig1}
\end{figure}

In figure \ref{number_intensity_fig1}, the dashed stepwise line represents the observational data and
the solid stepwise line represents the data corrected
for observational efficiencies which are derived in $\S$ \ref{observation}.
These cumulative number-intensity distributions show to be 
flattened below $4 \times 10^{-7}$ ergs cm$^{-2}$.
For the WXM instrument, the threshold sensitivity for localization is
of $3 \times 10^{-7}$ ergs cm$^{-2}$ in the 2-25 keV \citep{shi03}, 
which corresponds to $\sim 4 \times 10^{-7}$ ergs cm$^{-2}$ in 2-100
keV band, and hence the apparent flattening is thought to be due to 
the instrumental selection effect.

The cumulative distribution is well described by a single power law
model with a slope of $-1.4\pm0.4$ after correction for efficiency 
using the data above $4 \times 10^{-7}$\,ergs\,cm$^{-2}$.
Here the quoted error is for 90 \% confidence.
The fitting result is represented by the solid straight 
line in figure \ref{number_intensity_fig1}.
This slope is steeper than those obtained previously.
KONUS results gave a slope of $-0.9$ using 26 bursts with fluences above $\sim 10^{-6}$\,ergs\,cm$^{-2}$, 
which were observed between January 1979 and June 1999 \citep{apt01}.
The INTEGRAL result was $-0.91\pm0.09$ in the 15-100\,keV range using 224
bursts with fluences above $3 \times 10^{-8}$\,ergs\,cm$^{-2}$ during the period 
between March 2003 and October 2004 \citep{got06}.
BATSE results gave a somewhat flatter distribution with
$-0.76\pm0.17$ based on 92 bursts with fluences between $5.0 \times 10^{-8}$ and
$4.3 \times 10^{-6}$\,ergs\,cm$^{-2}$ observed between September 1993 
and June 1999 \citep{gog00}.
ICE results gave $-0.67\pm0.15$ using 113
bursts with fluences between $1.8 \times 10^{-7}$ and $6.5 \times 10^{-6}$\,ergs\,cm$^{-2}$ 
recorded between January 1979 and June 1984 \citep{gog00}.
The RXTE result was the flattest with a index of $-0.43\pm0.06$ 
for 266 bursts with fluences between $1.7 \times 10^{-10}$
and $1.9 \times 10^{-7}$\,ergs\,cm$^{-2}$ detected between 5 November 1996
and 18 November 1996 \citep{gog00}.

One possible explanation for these differences could be 
the different energy bands of the various experiments.
To investigate the energy dependence of the slope, we calculated
the cumulative distributions in two different bands, 2-10\,keV and 10-100\,keV. 
One can see the flattening below $6 \times 10^{-8}$\,ergs\,cm$^{-2}$
for 2-10\,keV and $3 \times 10^{-7}$\,ergs\,cm$^{-2}$ for 10-100\,keV.
The threshold sensitivity for localization corresponds to
$5.0 \times 10^{-8}$\,ergs\,cm$^{-2}$ for 2-10\,keV and
$3.0 \times 10^{-7}$\,ergs\,cm$^{-2}$ for 10-100\,keV \citep{shi03}. 
Thus we have restricted the fits to fluences above $6 \times 10^{-8}$\,ergs\,cm$^{-2}$
for 2-10\,keV and $3 \times 10^{-7}$\,ergs\,cm$^{-2}$ for 10-100\,keV.
In both energy bands, the distributions are well described by a single power law model.
The best-fit slopes are $-1.6\pm0.5$ for 2-10\,keV and $-1.4\pm0.4$ for 10-100\,keV, 
where the quoted errors again correspond to 90 \% confidence.
Thus, while there could be a slight energy dependence, it is not 
significant and does not fully explain these differences.

Another possibility is the intrinsic activity of SGRs, which would 
have been sampled differently by the previous experiments,
since they took data over different periods.
Although the observing period of HETE-2 overlapped with that of  INTEGRAL,
the HETE-2 bursts were distributed in June and July, while INTEGRAL recorded
bursts mostly in August, September and October \citep{got06}.

HETE-2 detected many bursts in the summer of 2004 when the source was very active.
Thus we investigated the cumulative number-intensity distribution using
only 2004 data. The dot-dashed stepwise line in figure
\ref{number_intensity_fig1} shows the corrected distribution and the
dot-dashed straight line shows a power law fit, which has a flatter slope of $-1.1\pm0.6$. 
This slope is consistent with those from KONUS and INTEGRAL.
This implies that more energetic bursts should occur relatively more frequently 
in periods of greater activity.

Since the slope for SGR\,1806$-$20 has been reported to be similar to that 
for SGR\,1900$+$14
\citep{apt01}, we compared the corrected distribution with the fluence of
an intermediate flare from SGR\,1900$+$14 detected by HETE-2.
The amount observing durations in each year are described in the parentheses of
table \ref{corrected_flares}.
Concerning the intermediate flare from SGR\,1900$+$14 detected by HETE-2,
the amount observing duration is 10450 ks from 2001 to 2005 and therefore
the burst rate is $8.3\times10^{-3}$ bursts day$^{-1}$,
where the quoted error corresponds to the Poisson error.
The 2-100\,keV fluence is estimated to be $1.9 \times 10^{-5}$\,ergs\,cm$^{-2}$ 
using the spectral parameters in \citet{oli04}.
It is shown as a star symbol in figure \ref{number_intensity_fig1}.
The fluence of the intermediate flare appears to be
fully consistent with the distribution of the short bursts in 2004, at $\sim 10^{-5}$\,ergs\,cm$^{-2}$.
This could imply a common origin or production mechanism for both 
short bursts and intermediate flares.
Considering the good fit with a single power law model,
it may also imply that there is no characteristic size for the short
bursts from SGR\,1806$-$20 up to $F \sim 10^{-5}$\,ergs\,cm$^{-2}$.

On the other hand, the fluence of the most energetic giant flare, from SGR\,1806$-$20 on December 27 2004 
is $\sim 2$\,ergs\,cm$^{-2}$ for the initial 0.6\,s \citep{ter05}
and the burst rate is $9.8\times10^{-5}$ bursts day$^{-1}$ taking into
account the time since the discovery of SGR\,1806$-$20 ($\sim28$ years).
This is clearly inconsistent with the distribution of the short bursts
and the intermediate flare (the difference in the energy bands has little effect).
This is consistent with the fact that giant flares are thought to
originate from different physical processes.

Although the fluence of the unusual flare from SGR\,1900$+$14
was $\sim 1.9 \times 10^{-5}$\,ergs\,cm$^{-2}$ \citep{ibr01}
which was same with the fluence of the intermediate flare from SGR\,1900$+$14,
its much longer duration of $\sim 1000$\,s was clearly inconsistent with the duration distribution.
This may imply that the taxonomy (i.e. short, intermediate, unusual or giant) is defined by
neither the duration nor the fluence.

The power law cumulative number-intensity distribution of SGR bursts is similar to that of
the earthquakes (e.g., \cite{kag99}) or the solar flares (e.g.,
\cite{den85});
the former relation is sometimes referred to as the Gutenberg-Richter law.
These imply that the SGR bursts could be due to the starquakes
or a similar process to the solar flares.
Considering that the slopes for the earthquakes are thought to be
influenced by, for example, such as the plate convergence rate (e.g., \cite{kag99}),
the difference of the slopes for SGR bursts might reflect
the intrinsic activities of the SGRs; in other words,
the starquakes in different zones of the neutron star surface would give the different slopes.

\subsection{Temporal Properties}\label{results_temporal_analyses}
Figure \ref{bright_single_flares} shows the light curves with 0.5\,ms and 5\,ms
time bins for six bright bursts.
These are classified as ``single'' peaked bursts by the procedure
using 5\,ms time bin data (see $\S$ \ref{observation}).
However, more complex and spiky structures are evident in the light curves with 0.5\,ms time bins.
Clearly the classification between single and multiple peaked bursts
is not based only on the intrinsic nature of the bursts themselves,
but is also highly dependent on the time resolution and/or statistics 
of the observations.

We find a delay of the softer emission compared with that 
in the 30-100\,keV band in the short bursts from SGR\,1806$-$20; 
$T_{\mathrm{lag}} = 2.2\pm0.4$\,ms for 2-10\,keV, and $T_{\mathrm{lag}} = 1.2\pm0.3$\,ms for 6-30\,keV.
Unfortunately the number of bursts is too small to estimate $T_{\mathrm{lag}}$ for SGR\,1900$+$14.
One possible explanation is rapid spectral softening.
Considering that the samples do not show a clear spectral evolution with a 20\,ms time resolution,
the spectral softening with a much faster time scale (a few milliseconds) should be required.
The cooling time scale of the emission from the higher blackbody component of 2BB is
${\tau}_{bb} = 0.08{\left(kT_{\mathrm{HT}}/11\,{\mathrm{keV}}\right)^{4}}{\left(R_{\mathrm{HT}}/4.5\,{\mathrm{km}}\right)^{3}}{\left(L_{\mathrm{HT}}/10^{40}\,{\mathrm{ergs\,s^{-1}}}\right)^{-1}}$\,ms,
where $L_{\mathrm{HT}}$ is a luminosity of the emission from the higher blackbody component of 2BB.
Assuming that $kT_{\mathrm{HT}}\sim$11\,keV, $R_{\mathrm{HT}}\sim$4.5\,km and $L_{\mathrm{HT}} \sim 10^{40}$\,ergs\,s$^{-1}$,
$\tau_{bb}$ turns out to be 0.08\,ms which is much smaller than 20\,ms.
Therefore the hypothesis of the very rapid spectral softening is plausible.
In addition, the time lag $T_{\mathrm{lag}}$ between the 2-5\,keV and 5-10\,keV time histories
is $T_{\mathrm{lag}} = 1.2\pm0.7$\,ms.
Although there is no enough statistics, the positive time lag may be due to the spectral softening.

\begin{figure}
 \begin{center}
  \FigureFile(80mm,79mm){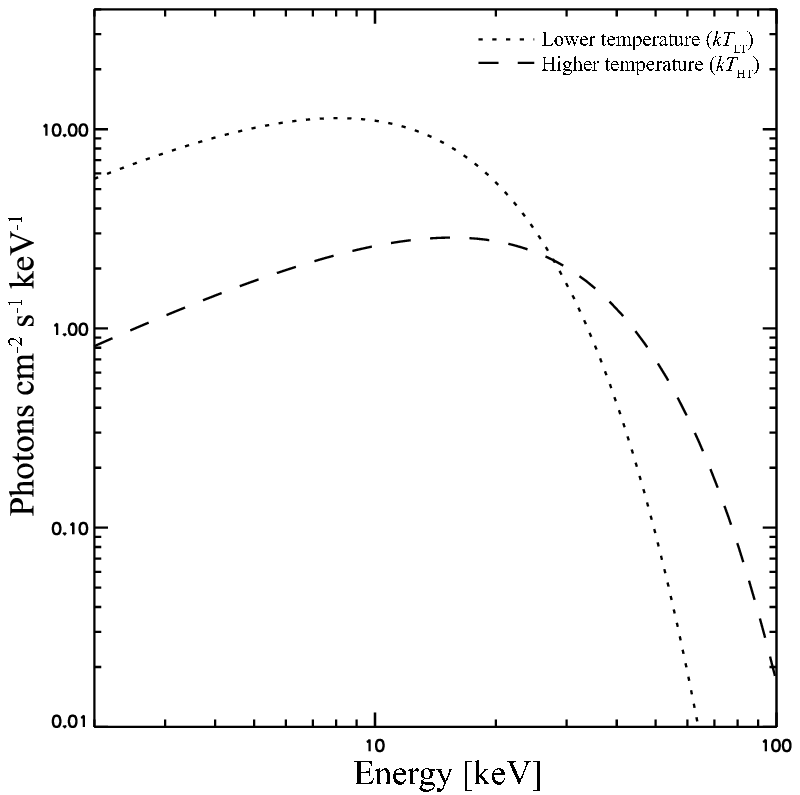}
 \end{center}
 \caption{Comparison of 2BB components for \#3387.}\label{compare_model_2bb}
\end{figure}

Alternative explanation would be the effect of separate emission regions.
Figure \ref{compare_model_2bb} shows the two spectral components of the 2BB model for \#3387.
We now evaluate which component dominates, and by how much, in the 2-10\,keV, 6-30\,keV,
and 30-100\,keV bands for this burst.
The ratios of counts expected from the lower temperature component to counts from the higher temperature component
are 5.3, 2.6 and 0.3 for 2-10\,keV, 6-30\,keV and 30-100\,keV, respectively.
Therefore the 2-10\,keV energy range represents the lower temperature component,
a nonzero $T_{\mathrm{lag}}$ between 2-10\,keV and 30-100\,keV
implies that these two components come from different regions and/or
different radiation mechanisms even though the 2BB model may only be empirical.
Thus the presence of the time lag supports
a multiple component model; at least, the spectra of short bursts 
consist of a softer and a harder emission component.

It is noteworthy that 
the small but clear time lag for SGR short bursts is different from the large time lags 
claimed for the long GRBs (e.g., \cite{nor02}).
Furthermore this is also different from the zero time lag for the short
GRBs (e.g., \cite{nor06}), while the short GRBs remain possible to be
generated from SGR giant flares in some scenarios (e.g., \cite{hur05}).

We find three bursts (indicated by $d$ in table \ref{flare_list1}) with
clear spectral softening, while three short bursts
(indicated by $e$ in table \ref{flare_list1}) might have a hard component later in the burst.
A possible origin for spectral softening in giant flares is the cooling of a
trapped fireball.
As we argue in $\S$ \ref{cumulative} and some theoretical works suggest 
\citep{dun94, lyu03}, the giant flares are presumably due to different
physical processes.
And hence, the trapped fireball does not seem appropriate to short bursts.
\citet{dun94} suggest that 
a small-scale crustal cracking of neutron star may trigger a short burst.
The crustal cracking causes the shift of the magnetic field footpoints.
The shift generates the Alfv\'{e}n waves which accelerate electrons.
Considering the short durations $\sim100$\,ms of the short bursts,
the accelerated electrons do not have to be trapped by a magnetic field,
or a fireball.
Consequently the accelerated electrons can simply lose their energies
and this may cause the spectral softening.
On the other hand, a possible emission mechanism for the hard component
could be Compton scattering by a hot plasma located at
most a few thousand kilometers away from the SGR  
(this estimate comes from the duration of the bursts).

A different scenario proposed by \citet{lyu03} is that 
the short bursts 
are due to heating of the magnetic corona caused by
local magnetic reconnections.
Moreover, the idea given by \citet{cea2006} suggests that
the glithces in the p-stars cause the energy injection into the magnetosphere 
and results in bursts.

\subsection{Spectral Properties}\label{spectral_properties}
\begin{figure}
 \begin{center}
  \FigureFile(80mm,79mm){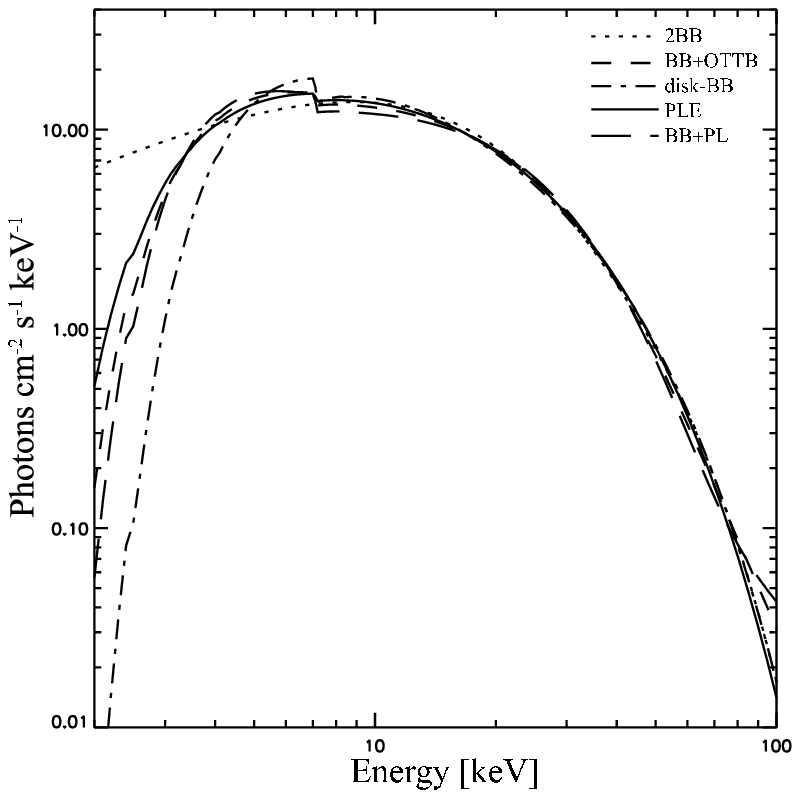}
 \end{center}
 \caption{Comparison of 5 fitting models for \#3387.}\label{compare_model}
\end{figure}

We tried eight types of spectral models and found  in $\S$ \ref{spectral_analyses} that
PLE, BB$+$OTTB, disk-BB, BB$+$PL and 2BB reproduce the observed spectra of the short bursts.
Figure \ref{compare_model} compares the spectral shapes of the best-fit models derived for \#3387.
Since there is no significant difference in their shapes between 5 and 80\,keV,
high quality data would be required to distinguish the models in these energy ranges.
The current burst data do not have enough statistics to distinguish between them.
However we note that there are signatures of slight excesses in the ranges below 5\,keV and/or above 80\,keV
if PLE, BB$+$OTTB, disk-BB and BB$+$PL are used,
while 2BB reproduces all the spectra in the samples quite well.

From the presence of the time lag, multiple component models such as BB$+$OTTB, BB$+$PL and 2BB are preferred.
This could suggest two separate emission regions, a hard component (30-100\,keV) region close to the SGR,
and a distant region ($\lesssim$700\,km) where the soft component (2-10\,keV) could be generated, presumably 
due to a delayed emission process.

If the SGRs are magnetars, the dominant radiation process in the vicinity of the SGRs should be synchrotron emission.
Assuming X-rays of $\sim10$\,keV are due to the synchrotron emission, 
the cooling time scale ${\tau}_s$ is evaluated as 
\[
  {\tau_{\mathrm{s}}} = {3 \times 10^{-15}}{\left(\frac{E_{\mathrm{e}}}{511\,\mathrm{keV}}\right)^{-1}}{\left(\frac{\mu}{5\times10^{32}\,\mathrm{G\,cm^{3}}}\right)^{-2}}{\left(\frac{R}{10^{7}\,\mathrm{cm}}\right)^{6}}\,\mathrm{s}
\]
where $E_{\mathrm{e}}$ is the electron energy, $\mu$ is the magnetic moment, and $R$ is
the distance from the center of the neutron star \citep{ryb79}.
Even for mildly relativistic electrons at $R \sim 100$\,km,
${\tau}_s$ turns out to be just $\sim3 \times 10^{-15}$\,s under the 
condition in which the magnetic field $B \sim 5 \times 10^{14}$\,G 
at the stellar surface. 
Therefore in the magnetar model, the power law type spectra in PLE or BB$+$PL might
be due to synchrotron emission from nonthermal electrons.

For PLE, the peak energy of the $\nu F_{\nu}$ spectrum $E_{\mathrm{p}}$ is
distributed mostly between 15 and 40\,keV.
Using the analogy of synchrotron shock models for GRBs, 
$E_{\mathrm{p}}$ may be associated with the electron energy distribution.
If this is the case, short bursts from SGRs could be related
to softer electrons than those required for hard GRB spectra
whose $E_{\mathrm{p}}$ distribution extends into the MeV region.
Also the very strong magnetic field environment might
give rise to non-linear magnetic QED effects between photons and the field.
Hard emission (over $\sim 1$\,MeV) could be suppressed
due to photon splitting, or soft X-rays could be enhanced 
by the lensing effect due to vacuum polarization \citep{koh02}.

The difficulty of this synchrotron-dominated PLE model is in 
the interpretation of the possible time lag in the single component framework,
and in explaining the large absorbing matter, necessary to get an acceptable fit,
which should be ejected prior to bursts.

In the case of BB$+$PL, the PL component may be explained as synchrotron emission.
Considering the steep PL indices, $\sim 2.1$, and the BB temperature,
$\sim$5.5\,keV, the soft emission is dominated by the PL component.
This appears to be inconsistent with the time lag,
which suggests that the soft X-rays are emitted from a distant location.
Furthermore some unknown radiation process must be invoked to dominate over synchrotron in 
the hard X-ray band, where the spectral shape can be reproduced by a
blackbody function with $kT\sim$5\,keV.

Fitting with BB$+$OTTB gives the best-fit temperatures 
$kT_{\mathrm{BB}}\sim$5\,keV and $kT_{\mathrm{OTTB}}\sim$27\,keV, and 
implies that the higher energy emission is due mainly to the OTTB component.
Considering the millisecond variations present in the short bursts,
the emission region should have a size $R \sim 300$\,km.
Therefore the electron density should be 
$n_{\mathrm{e}} \lesssim {\sigma_{\mathrm{T}}}^{-1} R^{-1} \sim 5 \times 10^{16}$\,cm$^{-3}$
for an optically thin region,  
where $\sigma_{\mathrm{T}}$ is the Thomson cross section.
OTTB is unlikely to be dominant if the SGR is a magnetar
because synchrotron emission should be more efficient 
than bremsstrahlung by many orders of magnitude ($B \sim 10^{15}$ G) 
for an optically thin plasma with $kT_{\mathrm{e}}\sim$10\,keV
and $n_{\mathrm{e}} \sim n_{\mathrm{p}} \sim 10^{16}$\,cm$^{-3}$.

The disk-BB model gives acceptable fits for most short bursts with a few exceptions.
This model is based on the physical picture of multiple blackbody emission coming from
an accretion disk heated by friction.
Most Galactic X-ray sources with disk-BB spectra are found in binary systems, but there is no evidence 
of companions for the SGR sources.
Alternatively, an accretion disk might be able to form, due to the fallback
of ejecta after the supernova explosion \citep{mic81, heg03}.
In any case, bright quiescent emission would be expected from the disk; 
however, this emission is dim for SGRs (about four orders of
magnitude smaller than that of short bursts).

Another possibility is that bursts come from an X-ray heated fallback disk \citep{per00}, 
which is heated not only by friction, but also by X-rays from the surface of the neutron star.
Recently, \citet{wan06} discovered such a disk around AXP\,4U\,0142$+$614.
If a similar disk is present around an SGR source, 
one would expect optical, infrared and submillimeter radiation 
from the irradiated disk, but no significant X-rays.
Much more energetic radiation is required to generate the X-ray emission observed in SGR bursts.
\citet{lyu03} proposed the heating of a magnetic corona around the neutron star
by local magnetic reconnection as the cause of short bursts.
If these high energy photons heat a disk, disk-BB type emission might be expected.
In this case, the inner radius of the irradiated fallback disk
$R_{\mathrm{in}}'$ \citep{per00} would be 
$R_{\mathrm{in}}' = 2.35 \times 10^{-2} 
        \left(T_{\mathrm{in}}'/{\mathrm{10^{4}\,eV}}\right)^{-7/3}
        \left(f/0.5\right)^{2/3}
        \eta^{-2/3}\left(L_{\mathrm{burst}}/10^{40}\,{\mathrm{ergs\,s^{-1}}}\right)^{2/3}
        \left(M/1.4\,\MO\right)^{-2/3}\,{\mathrm{km}}
$,
where $T_{\mathrm{in}}'$ is the temperature of the inner disk, $f$ is a factor
which expresses the uncertainty in the disk structure, 
$\eta$ is the efficiency of energy conversion from the irradiated energy
to the short burst energy, $L_{\mathrm{burst}}$ is the luminosity of the short burst,
$M$ is the mass of the neutron star and $\MO$ is the solar mass.
Assuming that $T_{\mathrm{in}}'\sim$10\,keV, $f\sim0.5$, $\eta\sim1$,
$L_{\mathrm{burst}} \sim 10^{40}$\,ergs\,s$^{-1}$ and
$M\sim$1.4\,$\MO$, $R_{\mathrm{in}}'$ turns out to be too small to be plausible, 0.024\,km.
Consequently an X-ray heated fallback disk is unlikely
as an origin for the short bursts, and no feasible physical
picture emerges from the disk-BB model. 

2BB fits represent all the short burst spectra well. 
\citet{oli04} suggest that a possible explanation of $kT_{\mathrm{HT}}$
for the intermediate flare is the emission from a trapped fireball.
However, it may be difficult to explain the short bursts by 
emission from a trapped fireball (see $\S$ \ref{results_temporal_analyses}).
As mentioned above, the dominant emission process should be synchrotron if a magnetar is the origin.
This suggests a possible inconsistency with the thermal emission.
The ``blackbody spectra'' in 2BB could simply be
apparent shapes, and 2BB would only be an empirical formula
to represent spectra of short bursts.

\begin{figure}
 \begin{center}
  \FigureFile(80mm,80mm){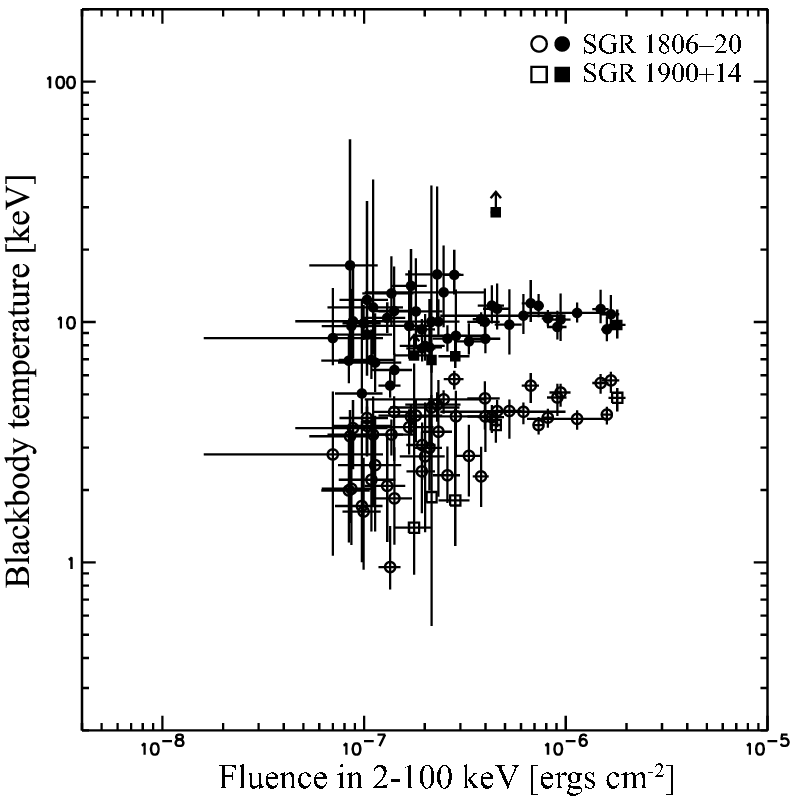}
 \end{center}
 \caption{Relation between the 2-100\,keV fluences and blackbody temperatures.
 The circle and filled circle show $kT_{\mathrm{LT}}$ and $kT_{\mathrm{HT}}$ respectively for SGR\,1806$-$20.
 The square and filled square show $kT_{\mathrm{LT}}$ and $kT_{\mathrm{HT}}$ respectively for SGR\,1900$+$14.}\label{fluence_kt}
\end{figure}

However it is remarkable that the two blackbody temperatures are 
found in clearly separated, rather narrow ranges  
around 4 and 11\,keV for all the bursts from  SGR\,1806$-$20 and
SGR\,1900$+$14, whose fluences are distributed over nearly 
two orders of magnitude.
Figure \ref{fluence_kt} shows the 2BB temperatures as a function of 2-100\,keV fluence.
These values, $\sim$4 and $\sim$11\,keV, are consistent with 
those found in previous studies which included much brighter events 
\citep{fer04, oli04, nak05, got06}.
The temperatures do not seem to depend on 
either the burst magnitude, the event morphology 
(i.e. single peaked or multiple peaked) or the source
(SGR\,1806$-$20 or SGR\,1900$+$14) (see also \cite{nak05}).

\begin{figure}
 \begin{center}
  \FigureFile(80mm,83mm){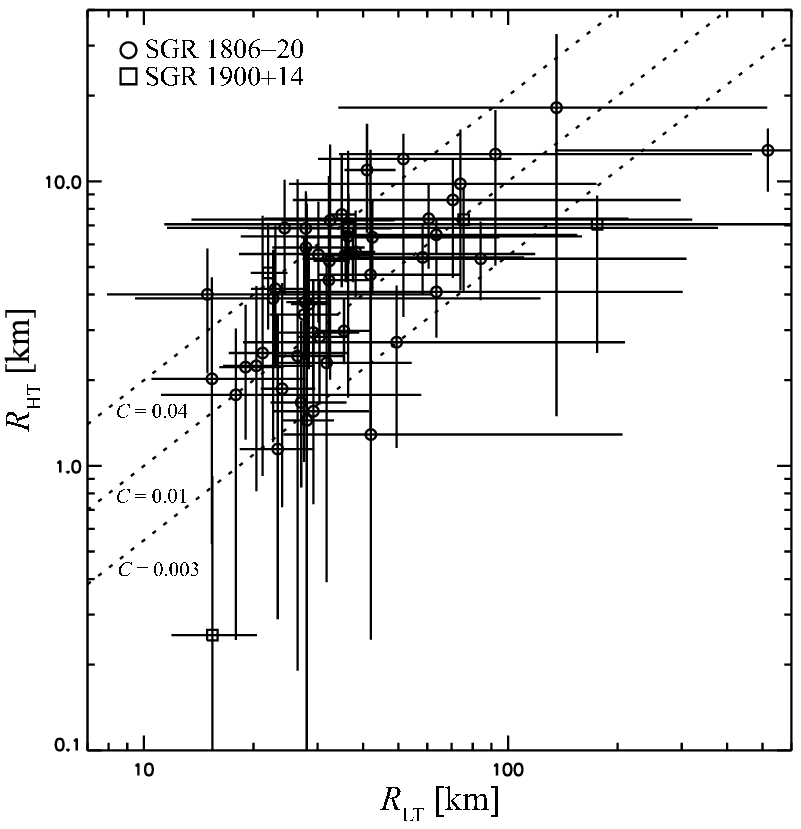}
 \end{center}
 \caption{Relation between the lower temperature blackbody emission radii ($R_{\mathrm{LT}}$)
 and those of the higher temperature blackbody ($R_{\mathrm{HT}}$).
 The dotted lines show that the ratio defined as $C = R_{\mathrm{HT}}^2/R_{\mathrm{LT}}^2$
 get constant value.
 They have a positive linear correlation with coefficient $r_{\mathrm{R}} = 0.79$.}\label{compare_radius}
\end{figure}

The (weighted) linear correlation coefficients between the fluence ${\mathrm{log}}(F)$
and ${\mathrm{log}}(kT_{\mathrm{LT}})$ or ${\mathrm{log}}(kT_{\mathrm{HT}})$ are $r_{\mathrm{LT}} = 0.66$ and
$r_{\mathrm{HT}} = 0.35$ respectively; $kT_{\mathrm{LT}}$
appears to have a slight positive correlation with fluence, while $kT_{\mathrm{HT}}$ does not.
The relation between $R_{\mathrm{LT}}$ and $R_{\mathrm{HT}}$ is shown in figure \ref{compare_radius}.
The dotted lines show that the ratio defined as $C = R_{\mathrm{HT}}^2/R_{\mathrm{LT}}^2$
get constant value.
The line with $C = 0.01$ seem to be consistent with the trend of the data.
This relation is theoretically proposed
in the framework of the p-stars \citep{cea2006}.
The (weighted) linear correlation coefficient between ${\mathrm{log}}(R_{\mathrm{LT}})$
and ${\mathrm{log}}(R_{\mathrm{HT}})$ is $r_{\mathrm{R}} = 0.79$.
These apparently constant temperatures may imply an unified view
of burst mechanisms and/or radiative transfer, for both SGR\,1806$-$20 and
SGR\,1900$+$14.

The other notable result is the time lag of 
the softer radiation with respect to the hard component,
$T_{\mathrm{lag}}\sim2.2$\,ms (see \ref{time_lag}).
The softer, $kT_{\mathrm{LT}}\sim$4\,keV, component could be 
reprocessed radiation from the harder, $kT_{\mathrm{HT}}\sim$11\,keV, 
emissions which might be generated near the SGR site (e.g. in the magnetosphere).
If this is the case, the distance to the reprocessing region from the $kT_{\mathrm{HT}}$ emitting
region is $d_{\mathrm{lag}} \lesssim 700$\,km,
and $R_{\mathrm{LT}} \sim 27$\,km could represent the size of
the reprocessing region (e.g. hot plasma or disk).

According to theoretical studies, a disk with an extremely high density of
$\rho\left(r_{\mathrm{A}}\right) \sim 10^{7}$\,g\,cm$^{-3}$ is formed around a neutron star
by fallback ejecta after a supernova explosion \citep{col71, mic81, heg03}.
The fallback disk could be a candidate of the reprocessing region which emitted
the softer, $kT_{\mathrm{LT}} \sim 4\,\mathrm{keV}$, emissions,
and should be located  outside the Alf\'{e}n radius.
Assuming a dipole magnetic field, the Alfv\'{e}n radius $r_{\mathrm{A}}$ is
evaluated as below.
\[
 r_{\mathrm{A}} = 14\left(\frac{\mu}{\mathrm{5\times10^{32}\,G\,cm^{3}}}\right)^{\frac{2}{5}}\left(\frac{M}{1.4\,\MO}\right)^{-\frac{1}{5}}{\left\{\frac{\rho\left(r_{\mathrm{A}}\right)}{\mathrm{10^{7}\,g\,cm^{-3}}}\right\}}^{-\frac{1}{5}}\,{\mathrm{km}}
\]
where $\mu$ is the magnetic moment which could be $\sim5 \times 10^{32}$\,G\,cm$^{3}$,
$M$ is the mass of the neutron star, $\MO$ is the solar mass 
and $\rho\left(r_{\mathrm{A}}\right)$ is the density of the medium.
As a result, $r_{\mathrm{A}}$ turns out to be 14\,km for $M$ = 1.4 $\MO$
even when the very dense disk is formed, 
and having weak dependence on the density
$r_{\mathrm{A}} \propto {\rho\left(r_{\mathrm{A}}\right)}^{-1/5}$.
Therefore the case that $d_{\mathrm{lag}} \gtrsim r_{\mathrm{A}}$ is plausible, 
and the reprocessed thermal emission could be generated 
from the hot plasma or a part of the disk.

For the PLE, BB$+$PL, BB$+$OTTB and disk-BB models, 
a large absorption ($\sim 10^{23}$\,cm$^{-2}$) is required to achieve acceptable fits.
The absorption for the quiescent emission is of the order of $10^{22}$\,cm$^{-2}$ and
therefore inconsistent with the large absorption during the short bursts.
Does a cool absorbing medium emerge only with the short bursts?
It is unlikely that cool matter ejection occurs 
together with or shortly before energetic emission from the 
source.
Fits with 2BB do not require this large absorption column, and
2BB is preferable for this reason, in addition to 
the presence of a time lag, even though it may be just a convenient empirical form which represents
the two component spectra.

Based on the discussions for temporal and spectral analyses
(see $\S$ \ref{results_temporal_analyses}
and $\S$ \ref{spectral_properties}),
two other explanations of emission mechanism of short bursts should be discussed.
The first explanation could be given by
very rapid (a few milliseconds) energy reinjection and
cooling at the emission region (see $\S$ \ref{results_temporal_analyses}).
In this framework, the nonzero time lag would be due to a spectral
softening with a time scale of a few milliseconds.
The harder component of 2BB would correspond to the energy reinjection
to the emission region (e.g. plasma in the magnetosphere),
while the softer component would correspond to its cooling phase.
As the second explanation, spatially diffused plasma trapped in the
magnetosphere could be heated and emit radiations; 
the Alfv\'{e}n waves due to the starquake, for example, thermalize the
contiguous regions of plasma from the inside to the outside 
along the magnetic field lines.
The Alfv\'{e}n velocity becomes very close to the velocity of light near
the surface of magnetars,
the propagating distance of the Alfv\'{e}n waves in $\sim2.2$\,ms
should be $\lesssim$700\,km.
Considering that it should be limited by
the distance along the magnetic field lines around the magnetars,
the emission radius of softer component $R_{\mathrm{LT}}\sim$27\,km is comparable.
In this case, the spectral shape should be reproduced by a 
multiple blackbody model and the 2BB spectra 
could just mimic its shape.

\section{Conclusion}
In five years from 2001 to 2005, HETE-2
localized 62 and 6 events from SGR\,1806$-$20 and SGR\,1900$+$14, respectively.
Its wide energy range, 2-400\,keV, allowed ideal studies
of the temporal and spectral properties of these events.
Among them, most bursts do not display any clear spectral 
(hardness) evolution with  rather long time scale of $\gtrsim 20$\,ms, 
with six exceptions among which three display clear spectral softening
and three others  might have a hard component at the end.

The data suggest a possible time lag $T_{\mathrm{lag}}\sim2.2$\,ms
for the softer radiation (2-10\,keV), compared to the harder emission (30-100\,keV).
One possible explanation could be by the {\it very rapid} 
spectral change, softening followed by very rapid re-hardening,  
with a time scale shorter than a few milliseconds, 
which were {\it not obseervable} by HETE-2 
because of the instrumental limit on time resolution.
This requires an emission mechanism
with very rapid ($\lesssim$ a few milliseconds) energy reinjection and cooling
at the emitting region, or the spatially contiguous heating of the
plasma with a nonuniform density in the magnetosphere near magnetars.

Alternatively, the nonzero time lag favors two emission regions.
The spectra of all short bursts from SGR\,1806$-$20 and SGR\,1900$+$14
are well fitted by 2BB with $kT_{\mathrm{LT}}\sim$4\,keV and
$kT_{\mathrm{HT}}\sim$11\,keV.
Considering the time lag, the X-rays corresponding 
to the  $kT_{\mathrm{HT}}$ ``blackbody'' may be emitted near the
SGRs and reprocessed by media located at $d_{\mathrm{lag}} \lesssim 700$\,km
into a softer ``blackbody'' component with $kT_{\mathrm{LT}}$.

The 2-100\,keV cumulative number-intensity distribution 
is represented by a power law with index $-1.4\pm0.4$
for all bursts, and $-1.1\pm0.6$ for bursts recorded only in 2004.
The fluences of the intermediate flare \citep{oli04} and the unusual flare
\citep{ibr01} from SGR\,1900$+$14 seem consistent with the
distribution for the 2004 data.
This may imply that more energetic bursts occur relatively more frequently 
in an active phase.

%%%%%%%%%%%%%%%%%%%%%%%%%%%%%%%%%%%%%%
% Acknowledgements
\bigskip
We would like to thank the members of the HETE-2 team for their support.
We would like to thank the anonymous referee for comments and
suggestions which improved our paper.
The HETE-2 mission is supported in the US by NASA contract NASW-4690, in
Japan in part by the Ministry of Education, Culture, Sports, Science,
and Technology Grant-in-Aid 14079102, and in France by CNES contract
793-01-8479.
KH is grateful for support under MIT Subcontract MIT-SC-R--293291.
YEN is supported by the JSPS Research Fellowships for Young Scientists. 

%%%%%%%%%%%%%%%%%%%%%%%%%%%%%%%%%%%%%%%

%%%%%%%%%%%%%%%%%%%%%%%%%%%%%%%%%%%%%%%
% Fugure
%%%%%%%%%%%%%%%%%%%%%%%%%%%%%%%%%%%%%%%
\clearpage

\begin{figure}
 \begin{center}
  \FigureFile(160mm,169mm){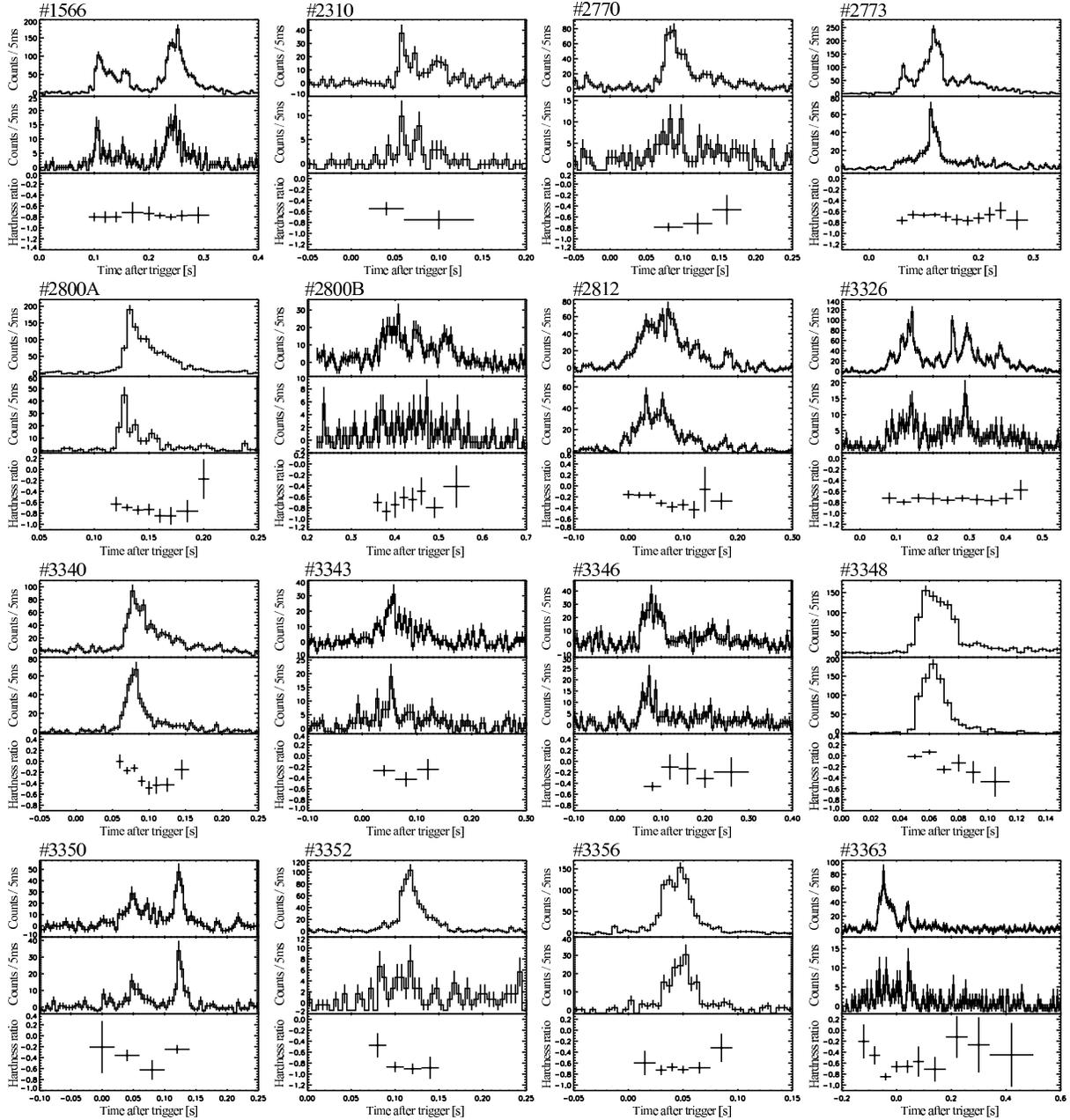}
 \end{center}
 \caption{Light curves observed by {\it FREGATE}
 in the 6-30\,keV ({\it top panels}) and the 30-100\,keV energy
 bands ({\it middle panels}), and time resolved hardness ratio
 ({\it bottom panels}) for all short bursts which do not lie on
 the dotted lines described in figure \ref{t90_dist}.
 The quoted errors are for 68 \% confidence.}\label{hr_plots1}
\end{figure}

\begin{figure}
 \begin{center}
  \FigureFile(160mm,86mm){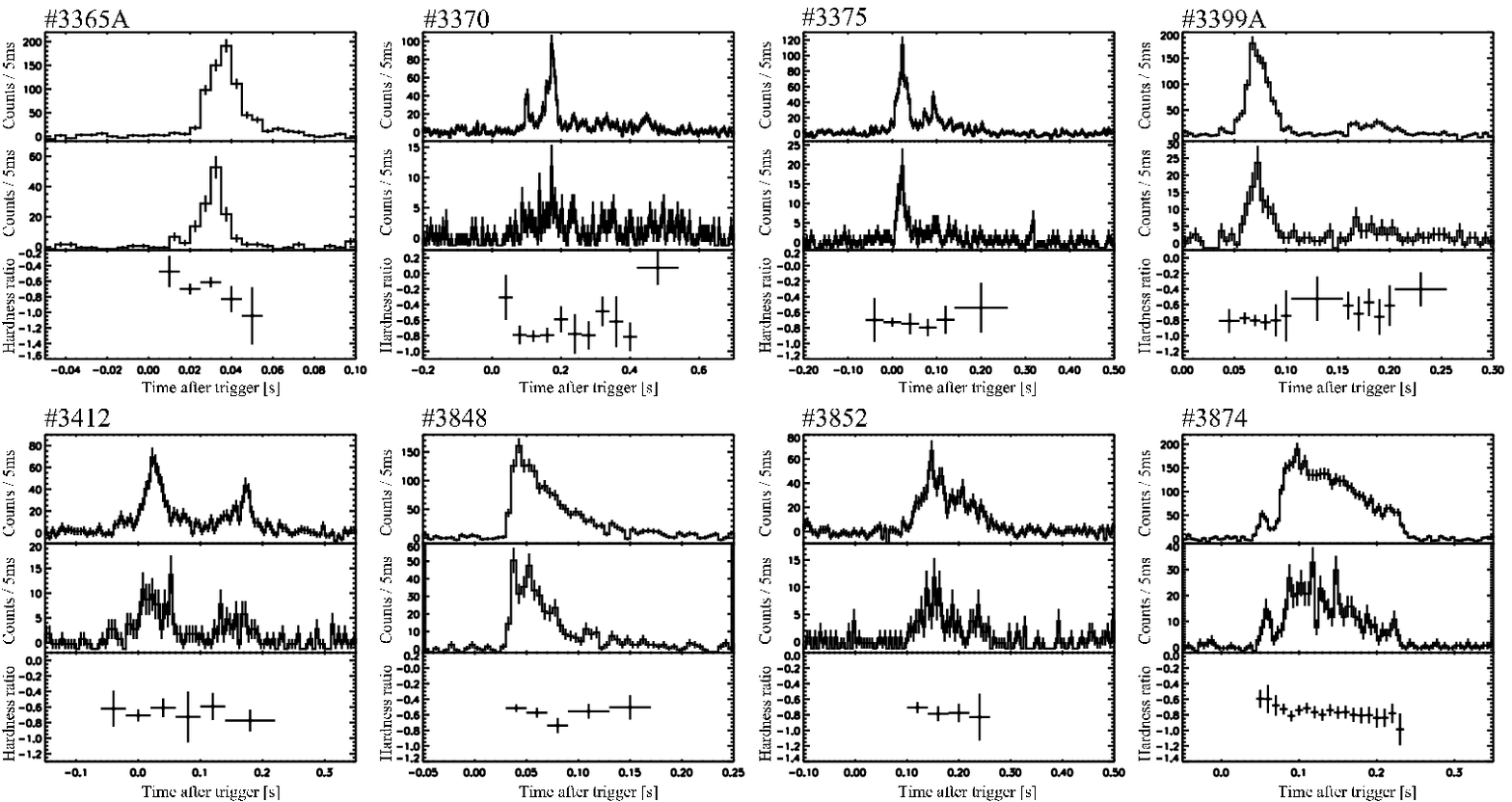}
 \end{center}
  \caption[4]{(Continued.)}
\end{figure}

%%%%%%%%%%%%%%%%%%%%%%%%%%%%%%%%%%%%%%%
% Table
%%%%%%%%%%%%%%%%%%%%%%%%%%%%%%%%%%%%%%%
\clearpage

\begin{table}
 \caption{The left column (``Localized Events'') for each SGR shows the number
 of localized events. The right columns (``Bursts'') are the number of bursts
 which we employ in our analyses for each SGR.
 Since one event contains two bursts in the case of \#2800, the number of events
 which we employ in our analyses is indicated in parentheses.
 The last (``Nonlocalized Events'') shows the number of nonlocalized
 events.}\label{evt_sum}
 \begin{center}
\begin{tabular}{ccrrcrrc}
   \hline\hline
   Year  & \multicolumn{3}{c}{SGR\,1806$-$20} & \multicolumn{3}{c}{SGR\,1900$+$14} & Nonlocalized Events \\
         & Localized Events & \multicolumn{2}{c}{Bursts} & Localized Events & \multicolumn{2}{c}{Bursts} & \\
   \hline
   2001  & 2   & 1  & (1)  & 5    & 4 & (4) & 5     \\
   2002  & 0   & 0  & (0)  & 1    & 1 & (1) & 5     \\
   2003  & 9   & 8  & (8)  & 0    & 0 & (0) & 12    \\
   2004  & 40  & 33 & (32) & 0    & 0 & (0) & 87    \\
   2005  & 11  & 8  & (8)  & 0    & 0 & (0) & 4     \\
   Total & 62  & 50 & (49) & 6    & 5 & (5) & 113   \\
   \hline
  \end{tabular}
 \end{center}
\end{table}

\begin{table}
 \caption{Number of localized bursts corrected for the observational
 efficiency, by year and source. The observing
 durations are given in parentheses in units of kiloseconds.}\label{corrected_flares}
 \begin{center}
  \begin{tabular}{crrrr}
   \hline\hline
   Year  & \multicolumn{2}{c}{SGR\,1806$-$20} & \multicolumn{2}{c}{SGR\,1900$+$14} \\
   \hline
   2001  & 7   & (3290)     & 12 & (1160)      \\
   2002  & 0   & (5140)     & 2  & (2820)      \\
   2003  & 27  & (4390)     & 0  & (3280)      \\
   2004  & 92  & (3700)     & 0  & (2090)      \\
   2005  & 21  & (3100)     & 0  & (1100)      \\
   Total & 147 & (19620)    & 14 & (10450)     \\
   \hline
  \end{tabular}
 \end{center}
\end{table}

\begin{longtable}{lllllllll}
 \caption{Main properties of the short bursts from SGR\,1806$-$20.}\label{flare_list1}
 \hline\hline
 Trigger \#\footnotemark[$*$] & \multicolumn{1}{l}{Trigger Date (UT)} & \multicolumn{3}{l}{$T_{90}$\footnotemark[$\dagger$] Duration (ms)} & \multicolumn{3}{l}{Total Counts} & Note\footnotemark[$\ddagger$] \\
            &                                       & 2-25\,keV & 6-30\,keV & 30-100\,keV & 2-25\,keV & 6-30\,keV & 30-100\,keV & \\
 \hline
 \endfirsthead
 \hline\hline
 Trigger \#\footnotemark[$*$] & \multicolumn{1}{l}{Trigger Date (UT)} & \multicolumn{3}{l}{$T_{90}$\footnotemark[$\dagger$] Duration (ms)} & \multicolumn{3}{l}{Total Counts} & Note\footnotemark[$\ddagger$] \\
            &                                       & 2-25\,keV & 6-30\,keV & 30-100\,keV & 2-25\,keV & 6-30\,keV & 30-100\,keV & \\
 \hline
 \endhead
 \hline
 \multicolumn{9}{l}{\hbox to 0pt{\parbox{180mm}{
 \footnotesize
 \footnotemark[$*$] HETE-2 trigger number.
 \par\noindent
 \footnotemark[$\dagger$] T90 is the time to accumulate between 5\% and 95\% of the counts.
 \par\noindent
 \footnotemark[$\ddagger$] Note for each burst;
 $a$ denotes a burst lying on the dotted lines described in figure \ref{t90_dist} for which we calculated the hardness ratio,%\\
 $b$ denotes single peaked bursts lying off the dotted line described in figure \ref{t90_dist},
 $c$ denotes multiple peaked bursts lying off the dotted line described in figure \ref{t90_dist},
 $d$ denotes bursts which have clear spectral softening,
 $e$ denotes bursts which have a hard component later in the burst,
 $h$ denotes bursts observed with the high gain configuration,
 $l$ denotes bursts which we employed to estimate $T_{\mathrm{lag}}$,
 $s$ denotes single peaked bursts,
 and $m$ denotes multiple peaked bursts.
 }}}
 \endfoot
 \hline
 \multicolumn{9}{l}{\hbox to 0pt{\parbox{180mm}{
 \footnotesize
 \footnotemark[$*$] HETE-2 trigger number.
 \par\noindent
 \footnotemark[$\dagger$] T90 is the time to accumulate between 5\% and 95\% of the counts.
 \par\noindent
 \footnotemark[$\ddagger$] Note for each burst;
 $a$ denotes a burst lying on the dotted lines described in figure \ref{t90_dist} for which we calculated the hardness ratio,%\\
 $b$ denotes single peaked bursts lying off the dotted line described in figure \ref{t90_dist},
 $c$ denotes multiple peaked bursts lying off the dotted line described in figure \ref{t90_dist},
 $d$ denotes bursts which have clear spectral softening,
 $e$ denotes bursts which have a hard component later in the burst,
 $h$ denotes bursts observed with the high gain configuration,
 $l$ denotes bursts which we employed to estimate $T_{\mathrm{lag}}$,
 $s$ denotes single peaked bursts,
 and $m$ denotes multiple peaked bursts.
 }}}
 \endlastfoot
 1566 & 2001:06:23.15:54:53 & 184$\pm$6 & 192$\pm$4 & 206$\pm$22 & 1020 & 2200 & 280 & $c$, $l$, $m$ \\
 2770 & 2003:07:18.23:43:12 & 136$\pm$37 & 116$\pm$13 & 196$\pm$42 & 363 & 660 & 125 & $b$, $s$ \\
 2771 & 2003:07:19.03:24:15 & 268$\pm$17 & 358$\pm$103 & 366$\pm$62 & 276 & 617 & 119 & $s$ \\
 2773 & 2003:07:19.22:34:15 & 196$\pm$29 & 188$\pm$5 & 234$\pm$43 & 957 & 2870 & 534 & $c$, $l$, $s$ \\
 2800A & 2003:08:11.16:06:28 & 72$\pm$21 & 68$\pm$8 & 106$\pm$40 & 441 & 1070 & 180 & $b$, $e$, $l$, $s$ \\
 2800B & 2003:08:11.16:06:28 & 160$\pm$27 & 172$\pm$15 & 198$\pm$29 & 254 & 461 & 86 & $b$, $s$ \\
 2802 & 2003:08:13.09:52:03 & 146$\pm$11 & 146$\pm$6 & 136$\pm$11 & 809 & 2020 & 2020 & $h$, $l$, $m$ \\
 2806 & 2003:08:15.11:15:24 & 98$\pm$28 & 96$\pm$40 & 84$\pm$19 & 126 & 293 & 194 & $h$, $s$ \\
\end{longtable}

\begin{longtable}{lllllllll}
 \caption{Main properties of the short bursts from SGR\,1900$+$14.}\label{flare_list2}
 \hline\hline
 Trigger \#\footnotemark[$*$] & \multicolumn{1}{l}{Trigger Date (UT)} & \multicolumn{3}{l}{$T_{90}$\footnotemark[$\dagger$] Duration (ms)} & \multicolumn{3}{l}{Total Counts} & Note\footnotemark[$\ddagger$] \\
            &                                       & 2-25\,keV & 6-30\,keV & 30-100\,keV & 2-25\,keV & 6-30\,keV & 30-100\,keV & \\
 \hline
 \endfirsthead
 \hline\hline
 Trigger \#\footnotemark[$*$] & \multicolumn{1}{l}{Trigger Date (UT)} & \multicolumn{3}{l}{$T_{90}$\footnotemark[$\dagger$] Duration (ms)} & \multicolumn{3}{l}{Total Counts} & Note\footnotemark[$\ddagger$] \\
            &                                       & 2-25\,keV & 6-30\,keV & 30-100\,keV & 2-25\,keV & 6-30\,keV & 30-100\,keV & \\
 \hline
 \endhead
 \hline
 \multicolumn{9}{l}{\hbox to 0pt{\parbox{180mm}{
 \footnotesize
 \footnotemark[$*$] HETE-2 trigger number.
 \par\noindent
 \footnotemark[$\dagger$] T90 is the time to accumulate between 5\% and 95\% of the counts.
 \par\noindent
 \footnotemark[$\ddagger$] Note for each burst;
 $a$ denotes a burst lying on the dotted lines described in figure \ref{t90_dist} for which we calculated the hardness ratio,
 $b$ denotes single peaked bursts lying off the dotted line described in figure \ref{t90_dist},
 $s$ denotes single peaked bursts,
 and $m$ denotes multiple peaked bursts.
 }}}
 \endfoot
 \hline
 \multicolumn{9}{l}{\hbox to 0pt{\parbox{180mm}{
 \footnotesize
 \footnotemark[$*$] HETE-2 trigger number.
 \par\noindent
 \footnotemark[$\dagger$] T90 is the time to accumulate between 5\% and 95\% of the counts.
 \par\noindent
 \footnotemark[$\ddagger$] Note for each burst;
 $a$ denotes a burst lying on the dotted lines described in figure \ref{t90_dist} for which we calculated the hardness ratio,
 $b$ denotes single peaked bursts lying off the dotted line described in figure \ref{t90_dist},
 $s$ denotes single peaked bursts,
 and $m$ denotes multiple peaked bursts.
 }}}
 \endlastfoot
 1568 & 2001:06:27.23:24:18 & 94$\pm$7 & 132$\pm$31 & 144$\pm$34 & 151 & 938 & 153 & $m$ \\
 1571 & 2001:06:28.15:06:36 & 300$\pm$27 & 294$\pm$12 & 300$\pm$34 & 673 & 4520 & 754 & $a$, $m$ \\
 1572 & 2001:06:29.12:13:32 & 98$\pm$8 & 104$\pm$25 & 100$\pm$8 & 98.9 & 692 & 106 & $a$, $s$ \\
 1578 & 2001:07:03.06:55:04 & 118$\pm$27 & 130$\pm$5 & 130$\pm$30 & 145 & 783 & 114 & $s$ \\
 2310 & 2002:08:27.15:35:15 & 84$\pm$18 & 86$\pm$26 & 60$\pm$13 & 126 & 217 & 42 & $b$, $s$ \\
\end{longtable}

\begin{longtable}{rllllllll}
 \caption{Spectral parameters obtained using 2BB for \#3259.}\label{spc_result_3259}
 \hline\hline
 Time Region & $N_{\mathrm{H}}$\footnotemark[$*$] &
 $kT_{\mathrm{LT}}$\footnotemark[$\dagger$] & $R_{\mathrm{LT}}\left(\frac{d}{15\,{\mathrm{kpc}}}\right)$\footnotemark[$\ddagger$] &
 $kT_{\mathrm{HT}}$\footnotemark[$\dagger$] & $R_{\mathrm{HT}}\left(\frac{d}{15\,{\mathrm{kpc}}}\right)$\footnotemark[$\ddagger$] &
 $F_{\mathrm{X}}$\footnotemark[$\S$] & $F_{\gamma}$\footnotemark[$\S$] & $\chi^2 (d.o.f.)$ \\
 (s) & ($10^{22}$\,cm$^{-2}$) & (keV) & (km) & (keV) & (km) &      &      & \\
 \hline
 \endfirsthead
 \hline\hline
 Time Region & $N_{\mathrm{H}}$\footnotemark[$*$] &
 $kT_{\mathrm{LT}}$\footnotemark[$\dagger$] & $R_{\mathrm{LT}}\left(\frac{d}{15\,{\mathrm{kpc}}}\right)$\footnotemark[$\ddagger$] &
 $kT_{\mathrm{HT}}$\footnotemark[$\dagger$] & $R_{\mathrm{HT}}\left(\frac{d}{15\,{\mathrm{kpc}}}\right)$\footnotemark[$\ddagger$] &
 $F_{\mathrm{X}}$\footnotemark[$\S$] & $F_{\gamma}$\footnotemark[$\S$] & $\chi^2 (d.o.f.)$ \\
 (s) & ($10^{22}$\,cm$^{-2}$) & (keV) & (km) & (keV) & (km) &      &      & \\
 \hline
 \endhead
 \hline
 \endfoot
 \hline
 \multicolumn{9}{l}{\hbox to 0pt{\parbox{180mm}{
 \footnotesize
 \footnotemark[$*$] Photoelectric absorption with 90 \% confidence level uncertainties.
 \par\noindent
 \footnotemark[$\dagger$] Blackbody temperatures with 90 \% confidence level uncertainties.
 \par\noindent
 \footnotemark[$\ddagger$] Emission radii with 90 \% confidence level uncertainties; $d$ is the distance to the source. We have used $d = 15$\,kpc.
 \par\noindent
 \footnotemark[$\S$] X-ray (2-10\,keV) and gamma-ray (10-100\,keV) fluxes, 10$^{-8}$\,ergs\,cm$^{-2}$\,s$^{-1}$, with 68 \% confidence level uncertainties.
 }}}
 \endlastfoot
 $-0.010-0.065$ & $<49$ & $2.6_{-0.9}^{+1.1}$ & $63_{-30}^{+147}$ & $7.4_{-1.2}^{+2.1}$ & $8.4_{-3.9}^{+4.9}$ & $39_{-16}^{+16}$ & $128_{-32}^{+32}$ & 22 (26) \\
 $0.065-0.120$  & $<23$ & $1.5_{-0.7}^{+0.8}$ & $97_{-52}^{+168}$ & $6.0_{-1.1}^{+1.5}$ & $10.8_{-4.5}^{+4.6}$ & ... & ... & 22 (21) \\
 $0.120-0.215$  & $11_{-7}^{+12}$ & $4.2_{-0.8}^{+0.6}$ & $46_{-9}^{+16}$ & $9.1_{-2.2}^{+5.3}$ & $6.1_{-4.4}^{+8.7}$ & $65_{-8}^{+8}$ & $348_{-89}^{+89}$ & 41 (47) \\
 $0.215-0.275$  & $<9$ & $5.0_{-3.4}^{+0.6}$ & $39_{-5}^{+65}$ & $9.4_{-3.5}^{+6.7}$ & $6.1_{-4.9}^{+13.3}$ & $85_{-10}^{+10}$ & $499_{-173}^{+173}$ & 43 (47) \\
 $0.275-0.395$  & $26_{-25}^{+46}$ & $1.5_{-0.5}^{+2.7}$ & $158_{-125}^{+336}$ & $5.4_{-0.3}^{+0.6}$ & $17.9_{-5.7}^{+5.2}$ & $30_{-17}^{+17}$ & $109_{-21}^{+21}$ & 22 (29) \\
 $0.395-0.440$  & $<32$ & $4.3_{-0.7}^{+0.6}$ & $27_{-7}^{+18}$ & $>10.4$ & $0.2_{-0.1}^{+1.1}$ & $28_{-7}^{+7}$ & $135_{-48}^{+48}$ & 10 (15) \\
 $0.440-0.520$  & $<9$ & $4.5_{-0.8}^{+0.5}$ & $41_{-6}^{+6}$ & $10.3_{-3.2}^{+18.1}$ & $4.0_{-3.5}^{+8.3}$ & $71_{-8}^{+8}$ & $342_{-101}^{+101}$ & 35 (40) \\
\end{longtable}

\begin{longtable}{rllllllll}
 \caption{Spectral parameters obtained using 2BB for \#3303.}\label{spc_result_3303}
 \hline\hline
 Time Region & $N_{\mathrm{H}}$\footnotemark[$*$] &
 $kT_{\mathrm{LT}}$\footnotemark[$\dagger$] & $R_{\mathrm{LT}}\left(\frac{d}{15\,{\mathrm{kpc}}}\right)$\footnotemark[$\ddagger$] &
 $kT_{\mathrm{HT}}$\footnotemark[$\dagger$] & $R_{\mathrm{HT}}\left(\frac{d}{15\,{\mathrm{kpc}}}\right)$\footnotemark[$\ddagger$] &
 $F_{\mathrm{X}}$\footnotemark[$\S$] & $F_{\gamma}$\footnotemark[$\S$] & $\chi^2 (d.o.f.)$ \\
 (s) & ($10^{22}$\,cm$^{-2}$) & (keV) & (km) & (keV) & (km) &      &      & \\
 \hline
 \endfirsthead
 \hline\hline
 Time Region & $N_{\mathrm{H}}$\footnotemark[$*$] &
 $kT_{\mathrm{LT}}$\footnotemark[$\dagger$] & $R_{\mathrm{LT}}\left(\frac{d}{15\,{\mathrm{kpc}}}\right)$\footnotemark[$\ddagger$] &
 $kT_{\mathrm{HT}}$\footnotemark[$\dagger$] & $R_{\mathrm{HT}}\left(\frac{d}{15\,{\mathrm{kpc}}}\right)$\footnotemark[$\ddagger$] &
 $F_{\mathrm{X}}$\footnotemark[$\S$] & $F_{\gamma}$\footnotemark[$\S$] & $\chi^2 (d.o.f.)$ \\
 (s) & ($10^{22}$\,cm$^{-2}$) & (keV) & (km) & (keV) & (km) &      &      & \\
 \hline
 \endhead
 \hline
 \endfoot
 \hline
 \multicolumn{9}{l}{\hbox to 0pt{\parbox{180mm}{
 \footnotesize
 \footnotemark[$*$] Photoelectric absorption with 90 \% confidence level uncertainties.
 \par\noindent
 \footnotemark[$\dagger$] Blackbody temperatures with 90 \% confidence level uncertainties.
 \par\noindent
 \footnotemark[$\ddagger$] Emission radii with 90 \% confidence level uncertainties; $d$ is the distance to the source. We have used $d = 15$\,kpc.
 \par\noindent
 \footnotemark[$\S$] X-ray (2-10\,keV) and gamma-ray (10-100\,keV) fluxes, 10$^{-8}$\,ergs\,cm$^{-2}$\,s$^{-1}$, with 68 \% confidence level uncertainties.
 }}}
 \endlastfoot
 $0.130-0.215$ & $<54$ & $3.8_{-2.0}^{+1.4}$ & $19_{-8}^{+13}$ & $11.1_{-4.1}^{+13.5}$ & $2.2_{-1.8}^{+1.9}$ & $12_{-10}^{+10}$ & $60_{-38}^{+38}$ & 13 (11) \\
 $0.215-0.330$ & $<37$ & $2.2_{-1.0}^{+1.5}$ & $49_{-31}^{+180}$ & $8.4_{-1.4}^{+1.6}$ & $4.8_{-1.7}^{+2.3}$ & $13_{-9}^{+9}$ & $59_{-16}^{+16}$ & 28 (30) \\
 $0.330-0.450$ & $<9$ & $3.6_{-1.1}^{+0.6}$ & $36_{-7}^{+21}$ & $10.3_{-1.7}^{+2.2}$ & $5.0_{-2.0}^{+3.0}$ & $41_{-38}^{+51}$ & $200_{-172}^{+1010}$ & 42 (39) \\
 $0.450-0.620$ & $<15$ & $3.9_{-1.0}^{+0.7}$ & $27_{-6}^{+17}$ & $9.1_{-2.1}^{+3.8}$ & $4.3_{-2.7}^{+4.4}$ & $24_{-5}^{+5}$ & $115_{-39}^{+39}$ & 25 (34) \\
 $0.620-0.705$ & $<7$ & $4.4_{-0.5}^{+0.5}$ & $30_{-5}^{+7}$ & $15.0_{-3.3}^{+6.3}$ & $2.1_{-1.0}^{+1.5}$ & $39_{-8}^{+8}$ & $218_{-48}^{+48}$ & 49 (48) \\
 $0.705-0.775$ & $<24$ & $5.0_{-1.0}^{+0.7}$ & $24_{-5}^{+11}$ & $12.3_{-3.9}^{+14.3}$ & $2.7_{-2.1}^{+4.7}$ & $31_{-7}^{+7}$ & $213_{-112}^{+112}$ & 32 (35) \\
\end{longtable}

\begin{longtable}{llllllll}
 \caption{Spectral parameters obtained using BB$+$OTTB for SGR\,1806$-$20.}\label{spc_list_bb_ottb_1}
 \hline\hline
 Trigger \#\footnotemark[$*$] &
 $N_{\mathrm{H}}$\footnotemark[$\dagger$] &
 $kT_{\mathrm{BB}}$\footnotemark[$\ddagger$] & $R_{\mathrm{BB}}\left(\frac{d}{15\,{\mathrm{kpc}}}\right)$\footnotemark[$\S$] &
 $kT_{\mathrm{OTTB}}$\footnotemark[$\|$] &
 $F_{\mathrm{X}}$\footnotemark[$\#$]   & $F_{\gamma}$\footnotemark[$\#$] &
 $\chi^2 (d.o.f.)$ \\
 & (10$^{22}$ cm$^{-2}$) & (keV) & (km) & (keV) & & & \\
 \hline
 \endfirsthead
 \hline\hline
 Trigger \#\footnotemark[$*$] &
 $N_{\mathrm{H}}$\footnotemark[$\dagger$] &
 $kT_{\mathrm{BB}}$\footnotemark[$\ddagger$] & $R_{\mathrm{BB}}\left(\frac{d}{15\,{\mathrm{kpc}}}\right)$\footnotemark[$\S$] &
 $kT_{\mathrm{OTTB}}$\footnotemark[$\|$] &
 $F_{\mathrm{X}}$\footnotemark[$\#$]   & $F_{\gamma}$\footnotemark[$\#$] &
 $\chi^2 (d.o.f.)$ \\
 & (10$^{22}$ cm$^{-2}$) & (keV) & (km) & (keV) & & & \\
 \hline
 \endhead
 \hline
 \multicolumn{8}{l}{\hbox to 0pt{\parbox{180mm}{
 \footnotesize
 \footnotemark[$*$] HETE-2 trigger number.
 \par\noindent
 \footnotemark[$\dagger$] Photoelectric absorption with 90 \% confidence level uncertainties.
 \par\noindent
 \footnotemark[$\ddagger$] Blackbody temperature with 90 \% confidence level uncertainties.
 \par\noindent
 \footnotemark[$\S$] Emission radius with 90 \% confidence level uncertainties; $d$ is the distance to the source. We have used $d = 15$\,kpc.
 \par\noindent
 \footnotemark[$\|$] Plasma temperature with 90 \% confidence level uncertainties.
 \par\noindent
 \footnotemark[$\#$] X-ray (2-10\,keV) and gamma-ray (10-100\,keV) fluxes, 10$^{-8}$\,ergs\,cm$^{-2}$\,s$^{-1}$, with 68 \% confidence level uncertainties.
 }}}
 \endfoot
 \hline
 \multicolumn{8}{l}{\hbox to 0pt{\parbox{180mm}{
 \footnotesize
 \footnotemark[$*$] HETE-2 trigger number.
 \par\noindent
 \footnotemark[$\dagger$] Photoelectric absorption with 90 \% confidence level uncertainties.
 \par\noindent
 \footnotemark[$\ddagger$] Blackbody temperature with 90 \% confidence level uncertainties.
 \par\noindent
 \footnotemark[$\S$] Emission radius with 90 \% confidence level uncertainties; $d$ is the distance to the source. We have used $d = 15$\,kpc.
 \par\noindent
 \footnotemark[$\|$] Plasma temperature with 90 \% confidence level uncertainties.
 \par\noindent
 \footnotemark[$\#$] X-ray (2-10\,keV) and gamma-ray (10-100\,keV) fluxes, 10$^{-8}$\,ergs\,cm$^{-2}$\,s$^{-1}$, with 68 \% confidence level uncertainties.
 }}}
 \endlastfoot
 1566 & 9$_{-6}^{+8}$ & 5.1$_{-0.5}^{+0.9}$     & 19$_{-8}^{+6}$   & 24$_{-6}^{+12}$ & 45$_{-5}^{+5}$ & 202$_{-30}^{+30}$ & 11 (34) \\
 2770 & $<$18 & 4.0$_{-0.7}^{+0.5}$             & 22$_{-7}^{+6}$   & 41$_{-19}^{+160}$ & 25$_{-4}^{+4}$ & 94$_{-28}^{+28}$ & 23 (32) \\
 2771 & 8$_{-7}^{+11}$ & 3.2$_{-1.0}^{+0.7}$    & 27$_{-10}^{+20}$ & 38$_{-17}^{+63}$ & 20$_{-4}^{+4}$ & 65$_{-20}^{+20}$ & 24 (39) \\
 2773 & 31$_{-8}^{+11}$ & 10.7$_{-5.9}^{+2.1}$  & 3$_{-2}^{+2}$    & 18$_{-6}^{+8}$ & 43$_{-7}^{+7}$ & 262$_{-29}^{+29}$ & 53 (57) \\
 2800A & 12$_{-10}^{+18}$ & 4.7$_{-1.2}^{+1.1}$ & 12$_{-8}^{+5}$   & 32$_{-7}^{+29}$ & 15$_{-2}^{+2}$ & 79$_{-17}^{+17}$ & 14 (25) \\
 2800B & 22$_{-10}^{+16}$ & ...             & ...          & 16$_{-3}^{+4}$   & 15$_{-2}^{+2}$ & 50$_{-7}^{+7}$ & 22 (29) \\
 2802 & 16$_{-8}^{+12}$ & 6.9$_{-0.6}^{+0.9}$   & 13$_{-4}^{+4}$   & 36$_{-9}^{+16}$ & 37$_{-3}^{+3}$ & 343$_{-44}^{+44}$ & 53 (62) \\
 2806 & 17$_{-10}^{+19}$ & ...              & ...          & 19$_{-5}^{+6}$   & 24$_{-4}^{+3}$ & 86$_{-12}^{+7}$ & 13 (26) \\
\end{longtable}

\begin{longtable}{llllllll}
 \caption{Spectral parameters obtained using BB$+$OTTB for SGR\,1900$+$14.}\label{spc_list_bb_ottb_2}
 \hline\hline
 Trigger \#\footnotemark[$*$] &
 $N_{\mathrm{H}}$\footnotemark[$\dagger$] &
 $kT_{\mathrm{BB}}$\footnotemark[$\ddagger$] & $R_{\mathrm{BB}}\left(\frac{d}{15\,{\mathrm{kpc}}}\right)$\footnotemark[$\S$] &
 $kT_{\mathrm{OTTB}}$\footnotemark[$\|$] &
 $F_{\mathrm{X}}$\footnotemark[$\#$]   & $F_{\gamma}$\footnotemark[$\#$] &
 $\chi^2 (d.o.f.)$ \\
 & (10$^{22}$ cm$^{-2}$) & (keV) & (km) & (keV) & & & \\
 \hline
 \endfirsthead
 \hline\hline
 Trigger \#\footnotemark[$*$] &
 $N_{\mathrm{H}}$\footnotemark[$\dagger$] &
 $kT_{\mathrm{BB}}$\footnotemark[$\ddagger$] & $R_{\mathrm{BB}}\left(\frac{d}{15\,{\mathrm{kpc}}}\right)$\footnotemark[$\S$] &
 $kT_{\mathrm{OTTB}}$\footnotemark[$\|$] &
 $F_{\mathrm{X}}$\footnotemark[$\#$]   & $F_{\gamma}$\footnotemark[$\#$] &
 $\chi^2 (d.o.f.)$ \\
 & (10$^{22}$ cm$^{-2}$) & (keV) & (km) & (keV) & & & \\
 \hline
 \endhead
 \hline
 \multicolumn{8}{l}{\hbox to 0pt{\parbox{180mm}{
 \footnotesize
 \footnotemark[$*$] HETE-2 trigger number.
 \par\noindent
 \footnotemark[$\dagger$] Photoelectric absorption with 90 \% confidence level uncertainties.
 \par\noindent
 \footnotemark[$\ddagger$] Blackbody temperature with 90 \% confidence level uncertainties.
 \par\noindent
 \footnotemark[$\S$] Emission radius with 90 \% confidence level uncertainties; $d$ is the distance to the source. We have used $d = 15$\,kpc.
 \par\noindent
 \footnotemark[$\|$] Plasma temperature with 90 \% confidence level uncertainties.
 \par\noindent
 \footnotemark[$\#$] X-ray (2-10\,keV) and gamma-ray (10-100\,keV) fluxes, 10$^{-8}$\,ergs\,cm$^{-2}$\,s$^{-1}$, with 68 \% confidence level uncertainties.
 }}}
 \endfoot
 \hline
 \multicolumn{8}{l}{\hbox to 0pt{\parbox{180mm}{
 \footnotesize
 \footnotemark[$*$] HETE-2 trigger number.
 \par\noindent
 \footnotemark[$\dagger$] Photoelectric absorption with 90 \% confidence level uncertainties.
 \par\noindent
 \footnotemark[$\ddagger$] Blackbody temperature with 90 \% confidence level uncertainties.
 \par\noindent
 \footnotemark[$\S$] Emission radius with 90 \% confidence level uncertainties; $d$ is the distance to the source. We have used $d = 15$\,kpc.
 \par\noindent
 \footnotemark[$\|$] Plasma temperature with 90 \% confidence level uncertainties.
 \par\noindent
 \footnotemark[$\#$] X-ray (2-10\,keV) and gamma-ray (10-100\,keV) fluxes, 10$^{-8}$\,ergs\,cm$^{-2}$\,s$^{-1}$, with 68 \% confidence level uncertainties.
 }}}
 \endlastfoot
 1568 & 41$_{-18}^{+30}$ & ...             & ...        & 20.2$_{-3.4}^{+4.2}$ & 27$_{-5}^{+5}$ & 164$_{-18}^{+18}$ & 24 (26) \\
 1571 & 8$_{-5}^{+7}$    & 6.4$_{-0.4}^{+0.5}$ & 13$_{-2}^{+2}$ & 27.3$_{-4.8}^{+6.9}$ & 66$_{-5}^{+5}$ & 426$_{-31}^{+31}$ & 64 (58) \\
 1572 & 42$_{-26}^{+57}$ & ...             & ...        & 23.3$_{-5.9}^{+9.0}$ & 21$_{-7}^{+7}$ & 157$_{-28}^{+28}$ & 9 (13) \\
 1578 & $<$37            & 7.0$_{-0.8}^{+1.0}$ & 7$_{-2}^{+2}$  & 6.9$_{-6.5}^{+21.6}$ & 25$_{-16}^{+16}$ & 113$_{-25}^{+25}$ & 5 (11) \\
 2310 & 9$_{-8}^{+21}$   & ...             & ...        & 17.3$_{-5.9}^{+9.9}$ & 16$_{-2}^{+2}$ & 41$_{-7}^{+7}$ & 16 (18) \\
\end{longtable}

\begin{longtable}{llllllll}
 \caption{Spectral parameters obtained using PLE for SGR\,1806$-$20.}\label{spc_list_cpl_1}
 \hline\hline
 Trigger \#\footnotemark[$*$] & $N_{\mathrm{H}}$\footnotemark[$\dagger$] & $\alpha$\footnotemark[$\ddagger$] &
 $E_{\mathrm{0}}$\footnotemark[$\S$] & $E_{\mathrm{p}}$\footnotemark[$\|$] & $F_{\mathrm{X}}$\footnotemark[$\#$] & $F_{\gamma}$\footnotemark[$\#$] &
 $\chi^2 (d.o.f.)$ \\
 & (10$^{22}$\,cm$^{-2}$) & & (keV) & (keV) & & & \\
 \hline
 \endfirsthead
 \hline\hline
 Trigger \#\footnotemark[$*$] & $N_{\mathrm{H}}$\footnotemark[$\dagger$] & $\alpha$\footnotemark[$\ddagger$] &
 $E_{\mathrm{0}}$\footnotemark[$\S$] & $E_{\mathrm{p}}$\footnotemark[$\|$] & $F_{\mathrm{X}}$\footnotemark[$\#$] & $F_{\gamma}$\footnotemark[$\#$] &
 $\chi^2 (d.o.f.)$ \\
 & (10$^{22}$\,cm$^{-2}$) & & (keV) & (keV) & & & \\
 \hline
 \endhead
 \hline
 \multicolumn{8}{l}{\hbox to 0pt{\parbox{180mm}{
 \footnotesize
 \footnotemark[$*$] HETE-2 trigger number.
 \par\noindent
 \footnotemark[$\dagger$] Photoelectric absorption with 90 \% confidence level uncertainties.
 \par\noindent
 \footnotemark[$\ddagger$] Spectral index with 90 \% confidence level uncertainties.
 \par\noindent
 \footnotemark[$\S$] Cutoff energy with 90 \% confidence level uncertainties.
 \par\noindent
 \footnotemark[$\|$] Peak energy of the ${\nu}F_{\nu}$ spectrum.
 \par\noindent
 \footnotemark[$\#$] X-ray (2-10\,keV) and gamma-ray (10-100\,keV) fluxes, 10$^{-8}$\,ergs\,cm$^{-2}$\,s$^{-1}$, with 68 \% confidence level uncertainties.
 }}}
 \endfoot
 \hline
 \multicolumn{8}{l}{\hbox to 0pt{\parbox{180mm}{
 \footnotesize
 \footnotemark[$*$] HETE-2 trigger number.
 \par\noindent
 \footnotemark[$\dagger$] Photoelectric absorption with 90 \% confidence level uncertainties.
 \par\noindent
 \footnotemark[$\ddagger$] Spectral index with 90 \% confidence level uncertainties.
 \par\noindent
 \footnotemark[$\S$] Cutoff energy with 90 \% confidence level uncertainties.
 \par\noindent
 \footnotemark[$\|$] Peak energy of the ${\nu}F_{\nu}$ spectrum.
 \par\noindent
 \footnotemark[$\#$] X-ray (2-10\,keV) and gamma-ray (10-100\,keV) fluxes, 10$^{-8}$\,ergs\,cm$^{-2}$\,s$^{-1}$, with 68 \% confidence level uncertainties.
 }}}
 \endlastfoot
 1566 & 9$_{-6}^{+10}$   & 0.23$_{-0.41}^{+0.42}$ & 12$_{-2}^{+3}$ & 21$_{-4}^{+6}$  & 46$_{-9}^{+9}$ & 195$_{-35}^{+35}$ & 15.2 (35) \\
 2770 & $<$29            & 0.44$_{-0.80}^{+0.89}$ & 13$_{-5}^{+12}$ & 21$_{-7}^{+19}$ & 23$_{-9}^{+9}$ & 84$_{-35}^{+35}$ & 13.4 (17) \\
 2771 & 23$_{-12}^{+21}$ & 1.82$_{-0.70}^{+0.57}$ & $>$20           & $>$10 & 20$_{-10}^{+10}$ & 61$_{-33}^{+33}$ & 26.2 (40) \\
 2773 & 17$_{-7}^{+9}$   & 0.48$_{-0.26}^{+0.28}$ & 17$_{-2}^{+3}$  & 26$_{-4}^{+4}$ & 42$_{-8}^{+8}$ & 259$_{-47}^{+47}$ & 67.0 (63) \\
 2800A & 22$_{-17}^{+38}$ & 0.86$_{-0.64}^{+0.89}$ & 20$_{-6}^{+20}$ & 22$_{-7}^{+22}$ & 15$_{-7}^{+7}$ & 76$_{-36}^{+36}$ & 17.7 (26) \\
 2800B & 17$_{-15}^{+34}$ & 0.74$_{-1.10}^{+1.17}$ & 14$_{-6}^{+22}$ & 18$_{-8}^{+28}$ & 15$_{-7}^{+7}$ & 50$_{-33}^{+33}$ & 21.8 (28) \\
 2802 & 13$_{-8}^{+13}$  & $-$0.02$_{-0.34}^{+0.37}$ & 15$_{-2}^{+3}$ & 30$_{-5}^{+6}$ & 38$_{-11}^{+11}$ & 342$_{-84}^{+84}$ & 56.4 (63) \\
 2806 & $<$39            & 0.56$_{-1.09}^{+1.11}$ & 15$_{-6}^{+22}$ & 21$_{-9}^{+32}$ & 23$_{-13}^{+13}$ & 86$_{-62}^{+62}$ & 12.6 (25) \\
\end{longtable}

\begin{longtable}{llllllll}
 \caption{Spectral parameters obtained using PLE for SGR\,1900$+$14.}\label{spc_list_cpl_2}
 \hline\hline
 Trigger \#\footnotemark[$*$] & $N_{\mathrm{H}}$\footnotemark[$\dagger$] & $\alpha$\footnotemark[$\ddagger$] &
 $E_{\mathrm{0}}$\footnotemark[$\S$] & $E_{\mathrm{p}}$\footnotemark[$\|$] & $F_{\mathrm{X}}$\footnotemark[$\#$] & $F_{\gamma}$\footnotemark[$\#$] &
 $\chi^2 (d.o.f.)$ \\
 & (10$^{22}$\,cm$^{-2}$) & & (keV) & (keV) & & & \\
 \hline
 \endfirsthead
 \hline\hline
 Trigger \#\footnotemark[$*$] & $N_{\mathrm{H}}$\footnotemark[$\dagger$] & $\alpha$\footnotemark[$\ddagger$] &
 $E_{\mathrm{0}}$\footnotemark[$\S$] & $E_{\mathrm{p}}$\footnotemark[$\|$] & $F_{\mathrm{X}}$\footnotemark[$\#$] & $F_{\gamma}$\footnotemark[$\#$] &
 $\chi^2 (d.o.f.)$ \\
 & (10$^{22}$\,cm$^{-2}$) & & (keV) & (keV) & & & \\
 \hline
 \endhead
 \hline
 \multicolumn{8}{l}{\hbox to 0pt{\parbox{180mm}{
 \footnotesize
 \footnotemark[$*$] HETE-2 trigger number.
 \par\noindent
 \footnotemark[$\dagger$] Photoelectric absorption with 90 \% confidence level uncertainties.
 \par\noindent
 \footnotemark[$\ddagger$] Spectral index with 90 \% confidence level uncertainties.
 \par\noindent
 \footnotemark[$\S$] Cutoff energy with 90 \% confidence level uncertainties.
 \par\noindent
 \footnotemark[$\|$] Peak energy of the ${\nu}F_{\nu}$ spectrum.
 \par\noindent
 \footnotemark[$\#$] X-ray (2-10\,keV) and gamma-ray (10-100\,keV) fluxes, 10$^{-8}$\,ergs\,cm$^{-2}$\,s$^{-1}$, with 68 \% confidence level uncertainties.
 }}}
 \endfoot
 \hline
 \multicolumn{8}{l}{\hbox to 0pt{\parbox{180mm}{
 \footnotesize
 \footnotemark[$*$] HETE-2 trigger number.
 \par\noindent
 \footnotemark[$\dagger$] Photoelectric absorption with 90 \% confidence level uncertainties.
 \par\noindent
 \footnotemark[$\ddagger$] Spectral index with 90 \% confidence level uncertainties.
 \par\noindent
 \footnotemark[$\S$] Cutoff energy with 90 \% confidence level uncertainties.
 \par\noindent
 \footnotemark[$\|$] Peak energy of the ${\nu}F_{\nu}$ spectrum.
 \par\noindent
 \footnotemark[$\#$] X-ray (2-10\,keV) and gamma-ray (10-100\,keV) fluxes, 10$^{-8}$\,ergs\,cm$^{-2}$\,s$^{-1}$, with 68 \% confidence level uncertainties.
 }}}
 \endlastfoot
 1568 & $<$50 &  0.04$_{-0.93}^{+1.02}$ & 13$_{-4}^{+5}$ & 25$_{-8}^{+10}$  & 26$_{-17}^{+17}$ & 159$_{-110}^{+110}$ & 21 (25) \\
 1571 & $<$11 & $-$0.26$_{-0.26}^{+0.27}$ & 11$_{-1}^{+1}$ & 26$_{-3}^{+3}$  & 68$_{-7}^{+7}$ & 426$_{-51}^{+51}$ & 67 (59) \\
 1572 & $<$82 & $-$0.05$_{-1.19}^{+1.68}$ & 13$_{-5}^{+30}$ & 27$_{-11}^{+62}$ & 21$_{-17}^{+17}$ & 149$_{-122}^{+122}$ & 7 (12) \\
 1578 & $<$17 & $-$0.56$_{-0.50}^{+0.67}$ & 10$_{-2}^{+3}$ & 25$_{-5}^{+9}$  & 21$_{-21}^{+21}$ & 118$_{-117}^{+117}$ & 7 (12) \\
 2310 & $<$56 & $<$2.61                 & $>$7           & ...  & 16$_{-1}^{+47}$ & 38$_{-31}^{+31}$ & 16 (17) \\
\end{longtable}

\begin{longtable}{lllllllll}
 \caption{Spectral parameters obtained using 2BB for SGR\,1806$-$20.}\label{spc_list_2bb_1}
 \hline\hline
 Trigger \#\footnotemark[$*$] & $N_{\mathrm{H}}$\footnotemark[$\dagger$] &
 $kT_{\mathrm{LT}}$\footnotemark[$\ddagger$] & $R_{\mathrm{LT}}\left(\frac{d}{15\,{\mathrm{kpc}}}\right)$\footnotemark[$\S$] &
 $kT_{\mathrm{HT}}$\footnotemark[$\ddagger$] & $R_{\mathrm{HT}}\left(\frac{d}{15\,{\mathrm{kpc}}}\right)$\footnotemark[$\S$] &
 $F_{\mathrm{X}}$\footnotemark[$\|$] & $F_{\gamma}$\footnotemark[$\|$] & $\chi^2 (d.o.f.)$ \\
 & ($10^{22}$\,cm$^{-2}$) & (keV) & (km) & (keV) & (km) &      &      & \\
 \hline
 \endfirsthead
 \hline\hline
 Trigger \#\footnotemark[$*$] & $N_{\mathrm{H}}$\footnotemark[$\dagger$] &
 $kT_{\mathrm{LT}}$\footnotemark[$\ddagger$] & $R_{\mathrm{LT}}\left(\frac{d}{15\,{\mathrm{kpc}}}\right)$\footnotemark[$\S$] &
 $kT_{\mathrm{HT}}$\footnotemark[$\ddagger$] & $R_{\mathrm{HT}}\left(\frac{d}{15\,{\mathrm{kpc}}}\right)$\footnotemark[$\S$] &
 $F_{\mathrm{X}}$\footnotemark[$\|$] & $F_{\gamma}$\footnotemark[$\|$] & $\chi^2 (d.o.f.)$ \\
 & ($10^{22}$\,cm$^{-2}$) & (keV) & (km) & (keV) & (km) &      &      & \\
 \hline
 \endhead
 \hline
 \multicolumn{9}{l}{\hbox to 0pt{\parbox{180mm}{
 \footnotesize
 \footnotemark[$*$] HETE-2 trigger number.
 \par\noindent
 \footnotemark[$\dagger$] Photoelectric absorption with 90 \% confidence level uncertainties.
 \par\noindent
 \footnotemark[$\ddagger$] Blackbody temperatures with 90 \% confidence level uncertainties.
 \par\noindent
 \footnotemark[$\S$] Emission radii with 90 \% confidence level uncertainties; $d$ is the distance to the source. We have used $d = 15$\,kpc.
 \par\noindent
 \footnotemark[$\|$] X-ray (2-10\,keV) and gamma-ray (10-100\,keV) fluxes, 10$^{-8}$\,ergs\,cm$^{-2}$\,s$^{-1}$, with 68 \% confidence level uncertainties.
 }}}
 \endfoot
 \hline
 \multicolumn{9}{l}{\hbox to 0pt{\parbox{180mm}{
 \footnotesize
 \footnotemark[$*$] HETE-2 trigger number.
 \par\noindent
 \footnotemark[$\dagger$] Photoelectric absorption with 90 \% confidence level uncertainties.
 \par\noindent
 \footnotemark[$\ddagger$] Blackbody temperatures with 90 \% confidence level uncertainties.
 \par\noindent
 \footnotemark[$\S$] Emission radii with 90 \% confidence level uncertainties; $d$ is the distance to the source. We have used $d = 15$\,kpc.
 \par\noindent
 \footnotemark[$\|$] X-ray (2-10\,keV) and gamma-ray (10-100\,keV) fluxes, 10$^{-8}$\,ergs\,cm$^{-2}$\,s$^{-1}$, with 68 \% confidence level uncertainties.
 }}}
 \endlastfoot
 1566 & $<4$ & $4.3_{-1.0}^{+0.5}$ & $32_{-4}^{+10}$ & $9.7_{-2.4}^{+3.9}$ & $4.5_{-2.3}^{+5.9}$ & $44_{-7}^{+7}$ & $206_{-37}^{+37}$ & 18 (34) \\
 2770 & $<6$ & $3.7_{-0.8}^{+0.7}$ & $28_{-5}^{+7}$ & $9.6_{-2.6}^{+6.8}$ & $3.4_{-2.4}^{+2.9}$ & $23_{-7}^{+7}$ & $89_{-35}^{+35}$ & 10 (16) \\
 2771 & $<10$ & $3.4_{-0.6}^{+0.5}$ & $29_{-7}^{+12}$ & $13.1_{-3.6}^{+5.6}$ & $1.6_{-0.8}^{+1.5}$ & $19_{-4}^{+4}$ & $66_{-24}^{+24}$ & 25 (39) \\
 2773 & $5_{-3}^{+5}$ & $4.0_{-0.4}^{+0.4}$ & $36_{-4}^{+6}$ & $10.4_{-0.8}^{+0.9}$ & $5.5_{-1.1}^{+1.2}$ & $43_{-3}^{+1}$ & $271_{-20}^{+20}$ & 38 (57) \\
 2800A & $<9$ & $4.3_{-0.6}^{+0.5}$ & $19_{-3}^{+5}$ & $11.3_{-2.0}^{+3.1}$ & $2.2_{-1.0}^{+1.5}$ & $15_{-4}^{+4}$ & $80_{-14}^{+14}$ & 14 (25) \\
 2800B & $<36$ & $2.5_{-1.2}^{+1.5}$ & $36_{-18}^{+123}$ & $6.8_{-1.7}^{+4.1}$ & $6.4_{-4.7}^{+6.4}$ & $15_{-8}^{+8}$ & $48_{-20}^{+20}$ & 23 (27) \\
 2802 & $<6$ & $5.4_{-0.8}^{+0.7}$ & $23_{-3}^{+5}$ & $12.0_{-1.9}^{+3.0}$ & $4.2_{-1.9}^{+2.8}$ & $34_{-3}^{+3}$ & $349_{-37}^{+37}$ & 57 (62) \\
 2806 & $<48$ & $2.0_{-0.8}^{+1.6}$ & $71_{-45}^{+227}$ & $6.9_{-1.3}^{+1.9}$ & $8.6_{-4.0}^{+3.4}$ & $24_{-12}^{+12}$ & $81_{-26}^{+26}$ & 12 (24) \\
\end{longtable}

\begin{longtable}{lllllllll}
 \caption{Spectral parameters obtained using 2BB for SGR\,1900$+$14.}\label{spc_list_2bb_2}
 \hline\hline
 Trigger \#\footnotemark[$*$] & $N_{\mathrm{H}}$\footnotemark[$\dagger$] &
 $kT_{\mathrm{LT}}$\footnotemark[$\ddagger$] & $R_{\mathrm{LT}}\left(\frac{d}{15\,{\mathrm{kpc}}}\right)$\footnotemark[$\S$] &
 $kT_{\mathrm{HT}}$\footnotemark[$\ddagger$] & $R_{\mathrm{HT}}\left(\frac{d}{15\,{\mathrm{kpc}}}\right)$\footnotemark[$\S$] &
 $F_{\mathrm{X}}$\footnotemark[$\|$] & $F_{\gamma}$\footnotemark[$\|$] & $\chi^2 (d.o.f.)$ \\
 & ($10^{22}$\,cm$^{-2}$) & (keV) & (km) & (keV) & (km) &      &      & \\
 \hline
 \endfirsthead
 \hline\hline
 Trigger \#\footnotemark[$*$] & $N_{\mathrm{H}}$\footnotemark[$\dagger$] &
 $kT_{\mathrm{LT}}$\footnotemark[$\ddagger$] & $R_{\mathrm{LT}}\left(\frac{d}{15\,{\mathrm{kpc}}}\right)$\footnotemark[$\S$] &
 $kT_{\mathrm{HT}}$\footnotemark[$\ddagger$] & $R_{\mathrm{HT}}\left(\frac{d}{15\,{\mathrm{kpc}}}\right)$\footnotemark[$\S$] &
 $F_{\mathrm{X}}$\footnotemark[$\|$] & $F_{\gamma}$\footnotemark[$\|$] & $\chi^2 (d.o.f.)$ \\
 & ($10^{22}$\,cm$^{-2}$) & (keV) & (km) & (keV) & (km) &      &      & \\
 \hline
 \endhead
 \hline
 \multicolumn{9}{l}{\hbox to 0pt{\parbox{180mm}{
 \footnotesize
 \footnotemark[$*$] HETE-2 trigger number.
 \par\noindent
 \footnotemark[$\dagger$] Photoelectric absorption with 90 \% confidence level uncertainties.
 \par\noindent
 \footnotemark[$\ddagger$] Blackbody temperatures with 90 \% confidence level uncertainties.
 \par\noindent
 \footnotemark[$\S$] Emission radii with 90 \% confidence level uncertainties; $d$ is the distance to the source. We have used $d = 15$\,kpc.
 \par\noindent
 \footnotemark[$\|$] X-ray (2-10\,keV) and gamma-ray (10-100\,keV) fluxes, 10$^{-8}$\,ergs\,cm$^{-2}$\,s$^{-1}$, with 68 \% confidence level uncertainties.
 }}}
 \endfoot
 \hline
 \multicolumn{9}{l}{\hbox to 0pt{\parbox{180mm}{
 \footnotesize
 \footnotemark[$*$] HETE-2 trigger number.
 \par\noindent
 \footnotemark[$\dagger$] Photoelectric absorption with 90 \% confidence level uncertainties.
 \par\noindent
 \footnotemark[$\ddagger$] Blackbody temperatures with 90 \% confidence level uncertainties.
 \par\noindent
 \footnotemark[$\S$] Emission radii with 90 \% confidence level uncertainties; $d$ is the distance to the source. We have used $d = 15$\,kpc.
 \par\noindent
 \footnotemark[$\|$] X-ray (2-10\,keV) and gamma-ray (10-100\,keV) fluxes, 10$^{-8}$\,ergs\,cm$^{-2}$\,s$^{-1}$, with 68 \% confidence level uncertainties.
 }}}
 \endlastfoot
 1568 & $<169$ & $1.8_{-0.6}^{+2.6}$ & $76_{-62}^{+245}$ & $7.2_{-0.8}^{+1.5}$ & $7.3_{-3.2}^{+2.1}$ & $28_{-21}^{+21}$ & $149_{-23}^{+23}$ & 20 (24) \\
 1571 & $<2$ & $4.8_{-0.6}^{+0.5}$ & $22_{-2}^{+3}$ & $9.7_{-1.2}^{+1.5}$ & $4.8_{-1.8}^{+2.4}$ & $62_{-9}^{+9}$ & $438_{-47}^{+47}$ & 67 (58) \\
 1572 & $<93$ & $1.4_{-0.5}^{+5.3}$ & $176_{-164}^{+1737}$ & $>6.2$ & $7.1_{-4.6}^{+1.8}$ & $23_{-20}^{+20}$ & $135_{-30}^{+30}$ & 7 (11) \\
 1578 & $<30$ & $1.9_{-1.3}^{+1.9}$ & $36_{-25}^{+341}$ & $6.9_{-0.8}^{+0.8}$ & $6.9_{-1.7}^{+2.2}$ & ... & ... & 5 (11) \\
 2310 & $<11$ & $3.7_{-0.6}^{+0.5}$ & $15_{-4}^{+5}$ & $>12.5$ & $0.3_{-0.2}^{+0.7}$ & $16_{-6}^{+6}$ & $57_{-26}^{+26}$ & 8 (16) \\
\end{longtable}

\begin{longtable}{lllllll}
 \caption{Spectral parameters obtained using disk-BB for SGR\,1806$-$20.}\label{spc_list_diskbb_1}
 \hline\hline
 Trigger \#\footnotemark[$*$] & $N_{\mathrm{H}}$\footnotemark[$\dagger$] &
 $T_{\mathrm{in}}$\footnotemark[$\ddagger$] & $R_{\mathrm{in}}\sqrt{\cos\theta}\left(\frac{d}{15\,{\mathrm{kpc}}}\right)$\footnotemark[$\S$] &
 $F_{\mathrm{X}}$\footnotemark[$\|$] & $F_{\gamma}$\footnotemark[$\|$] & $\chi^2 (d.o.f.)$ \\
 & (10$^{22}$\,cm$^{-2}$) & (keV) & (km) & & & \\
 \hline
 \endfirsthead
 \hline\hline
 Trigger \#\footnotemark[$*$] & $N_{\mathrm{H}}$\footnotemark[$\dagger$] &
 $T_{\mathrm{in}}$\footnotemark[$\ddagger$] & $R_{\mathrm{in}}\sqrt{\cos\theta}\left(\frac{d}{15\,{\mathrm{kpc}}}\right)$\footnotemark[$\S$] &
 $F_{\mathrm{X}}$\footnotemark[$\|$] & $F_{\gamma}$\footnotemark[$\|$] & $\chi^2 (d.o.f.)$ \\
 & (10$^{22}$\,cm$^{-2}$) & (keV) & (km) & & & \\
 \hline
 \endhead
 \hline
 \multicolumn{7}{l}{\hbox to 0pt{\parbox{180mm}{
 \footnotesize
 \footnotemark[$*$] HETE-2 trigger number.
 \par\noindent
 \footnotemark[$\dagger$] Photoelectric absorption with 90 \% confidence level uncertainties.
 \par\noindent
 \footnotemark[$\ddagger$] Inner disk radius temperature with 90 \% confidence level uncertainties.
 \par\noindent
 \footnotemark[$\S$] $R_{\mathrm{in}}$ is the inner disk radius, $\theta$ is the angle of the disk, and $d$ is the distance to the source.\\
 We used $d = 15$\,kpc. The quoted uncertainties are for 90 \% confidence.
 \par\noindent
 \footnotemark[$\|$] X-ray (2-10\,keV) and gamma-ray (10-100\,keV) fluxes, 10$^{-8}$\,ergs\,cm$^{-2}$\,s$^{-1}$, with 68 \% confidence level uncertainties.
 }}}
 \endfoot
 \hline
 \multicolumn{7}{l}{\hbox to 0pt{\parbox{180mm}{
 \footnotesize
 \footnotemark[$*$] HETE-2 trigger number.
 \par\noindent
 \footnotemark[$\dagger$] Photoelectric absorption with 90 \% confidence level uncertainties.
 \par\noindent
 \footnotemark[$\ddagger$] Inner disk radius temperature with 90 \% confidence level uncertainties.
 \par\noindent
 \footnotemark[$\S$] $R_{\mathrm{in}}$ is the inner disk radius, $\theta$ is the angle of the disk, and $d$ is the distance to the source.\\
 We used $d = 15$\,kpc. The quoted uncertainties are for 90 \% confidence.
 \par\noindent
 \footnotemark[$\|$] X-ray (2-10\,keV) and gamma-ray (10-100\,keV) fluxes, 10$^{-8}$\,ergs\,cm$^{-2}$\,s$^{-1}$, with 68 \% confidence level uncertainties.
 }}}
 \endlastfoot
 1566 & 10$_{-4}^{+5}$  & 9.4$_{-0.6}^{+0.6}$ & 6.1$_{-0.8}^{+1.0}$ & 46$_{-4}^{+4}$ & 193$_{-11}^{+11}$ & 17 (36) \\
 2770 & 7$_{-5}^{+8}$   & 9.1$_{-1.1}^{+1.2}$ & 4.3$_{-1.0}^{+1.4}$ & 23$_{-3}^{+3}$ & 83$_{-12}^{+12}$ & 15 (18) \\
 2771 & 8$_{-5}^{+8}$   & 7.6$_{-1.2}^{+1.5}$ & 5.2$_{-1.7}^{+2.3}$ & 19$_{-3}^{+3}$ & 53$_{-10}^{+10}$ & 37 (41) \\
 2773 & 14$_{-3}^{+4}$  & 11.5$_{-0.6}^{+0.6}$ & 4.5$_{-0.5}^{+0.6}$ & 42$_{-2}^{+2}$ & 259$_{-10}^{+10}$ & 58 (59) \\
 2800A & 9$_{-5}^{+8}$  & 10.7$_{-1.0}^{+1.1}$ & 2.8$_{-0.5}^{+0.7}$ & 15$_{-2}^{+2}$ & 73$_{-6}^{+6}$ & 23 (27) \\
 2800B & 10$_{-7}^{+12}$ & 8.2$_{-1.2}^{+1.4}$ & 4.2$_{-1.2}^{+1.7}$ & 14$_{-3}^{+3}$ & 49$_{-11}^{+11}$ & 23 (29) \\
 2802 & 23$_{-7}^{+9}$  & 12.9$_{-0.8}^{+0.8}$ & 4.1$_{-0.5}^{+0.6}$ & 38$_{-3}^{+3}$ & 342$_{-16}^{+16}$ & 59 (64) \\
 2806 & 7$_{-7}^{+13}$  & 9.4$_{-1.6}^{+1.9}$ & 4.0$_{-1.3}^{+2.0}$ & 22$_{-5}^{+5}$ & 85$_{-25}^{+25}$ & 13 (26) \\
\end{longtable}

\begin{longtable}{lllllll}
 \caption{Spectral parameters obtained using disk-BB for SGR\,1900$+$14.}\label{spc_list_diskbb_2}
 \hline\hline
 Trigger \#\footnotemark[$*$] & $N_{\mathrm{H}}$\footnotemark[$\dagger$] &
 $T_{\mathrm{in}}$\footnotemark[$\ddagger$] & $R_{\mathrm{in}}\sqrt{\cos\theta}\left(\frac{d}{15\,{\mathrm{kpc}}}\right)$\footnotemark[$\S$] &
 $F_{\mathrm{X}}$\footnotemark[$\|$] & $F_{\gamma}$\footnotemark[$\|$] & $\chi^2 (d.o.f.)$ \\
 & (10$^{22}$\,cm$^{-2}$) & (keV) & (km) & & & \\
 \hline
 \endfirsthead
 \hline\hline
 Trigger \#\footnotemark[$*$] & $N_{\mathrm{H}}$\footnotemark[$\dagger$] &
 $T_{\mathrm{in}}$\footnotemark[$\ddagger$] & $R_{\mathrm{in}}\sqrt{\cos\theta}\left(\frac{d}{15\,{\mathrm{kpc}}}\right)$\footnotemark[$\S$] &
 $F_{\mathrm{X}}$\footnotemark[$\|$] & $F_{\gamma}$\footnotemark[$\|$] & $\chi^2 (d.o.f.)$ \\
 & (10$^{22}$\,cm$^{-2}$) & (keV) & (km) & & & \\
 \hline
 \endhead
 \hline
 \multicolumn{7}{l}{\hbox to 0pt{\parbox{180mm}{
 \footnotesize
 \footnotemark[$*$] HETE-2 trigger number.
 \par\noindent
 \footnotemark[$\dagger$] Photoelectric absorption with 90 \% confidence level uncertainties.
 \par\noindent
 \footnotemark[$\ddagger$] Inner disk radius temperature with 90 \% confidence level uncertainties.
 \par\noindent
 \footnotemark[$\S$] $R_{\mathrm{in}}$ is the inner disk radius, $\theta$ is the angle of the disk, and $d$ is the distance to the source.\\
 We used $d = 15$\,kpc. The quoted uncertainties are for 90 \% confidence.
 \par\noindent
 \footnotemark[$\|$] X-ray (2-10\,keV) and gamma-ray (10-100\,keV) fluxes, 10$^{-8}$\,ergs\,cm$^{-2}$\,s$^{-1}$, with 68 \% confidence level uncertainties.
 }}}
 \endfoot
 \hline
 \multicolumn{7}{l}{\hbox to 0pt{\parbox{180mm}{
 \footnotesize
 \footnotemark[$*$] HETE-2 trigger number.
 \par\noindent
 \footnotemark[$\dagger$] Photoelectric absorption with 90 \% confidence level uncertainties.
 \par\noindent
 \footnotemark[$\ddagger$] Inner disk radius temperature with 90 \% confidence level uncertainties.
 \par\noindent
 \footnotemark[$\S$] $R_{\mathrm{in}}$ is the inner disk radius, $\theta$ is the angle of the disk, and $d$ is the distance to the source.\\
 We used $d = 15$\,kpc. The quoted uncertainties are for 90 \% confidence.
 \par\noindent
 \footnotemark[$\|$] X-ray (2-10\,keV) and gamma-ray (10-100\,keV) fluxes, 10$^{-8}$\,ergs\,cm$^{-2}$\,s$^{-1}$, with 68 \% confidence level uncertainties.
 }}}
 \endlastfoot
 1568 & 18$_{-11}^{+21}$ & 10.5$_{-1.2}^{+1.4}$ & 2.9$_{-0.7}^{+0.9}$ & 26$_{-7}^{+7}$ & 158$_{-24}^{+24}$ & 21 (26) \\
 1571 & 17$_{-5}^{+7}$   & 10.8$_{-0.5}^{+0.5}$ & 4.4$_{-0.4}^{+0.5}$ & 69$_{-4}^{+4}$ & 427$_{-15}^{+15}$ & 75 (60) \\
 1572 & 20$_{-17}^{+41}$ & 11.4$_{-1.9}^{+2.4}$ & 2.4$_{-0.8}^{+1.2}$ & 21$_{-8}^{+8}$ & 149$_{-42}^{+42}$ & 7 (13) \\
 1578 & $<$46            & 10.2$_{-1.3}^{+1.4}$ & 2.6$_{-0.7}^{+1.1}$ & 23$_{-6}^{+6}$ & 117$_{-22}^{+22}$ & 10 (13) \\
 2310 & $<$23            & 7.1$_{-1.5}^{+2.1}$  & 3.4$_{-1.5}^{+2.5}$ & 16$_{-5}^{+5}$ & 38$_{-13}^{+13}$ & 16 (18) \\
\end{longtable}

\begin{longtable}{llllllll}
 \caption{Spectral parameters obtained using BB$+$PL for SGR\,1806$-$20.}\label{spc_list_bb_pow_1}
 \hline\hline
 Trigger \#\footnotemark[$*$] & $N_{\mathrm{H}}$\footnotemark[$\dagger$] &
 $kT_{\mathrm{BB}}$\footnotemark[$\ddagger$] & $R_{\mathrm{BB}}\left(\frac{d}{15\,{\mathrm{kpc}}}\right)$\footnotemark[$\S$] &
 $\alpha$\footnotemark[$\|$] & $F_{\mathrm{X}}$\footnotemark[$\#$] & $F_{\gamma}$\footnotemark[$\#$] & $\chi^2 (d.o.f.)$ \\
 & (10$^{22}$\,cm$^{-2}$) & (keV) & (km) & & & & \\
 \hline
 \endfirsthead
 \hline\hline
 Trigger \#\footnotemark[$*$] & $N_{\mathrm{H}}$\footnotemark[$\dagger$] &
 $kT_{\mathrm{BB}}$\footnotemark[$\ddagger$] & $R_{\mathrm{BB}}\left(\frac{d}{15\,{\mathrm{kpc}}}\right)$\footnotemark[$\S$] &
 $\alpha$\footnotemark[$\|$] & $F_{\mathrm{X}}$\footnotemark[$\#$] & $F_{\gamma}$\footnotemark[$\#$] & $\chi^2 (d.o.f.)$ \\
 & (10$^{22}$\,cm$^{-2}$) & (keV) & (km) & & & & \\
 \hline
 \endhead
 \hline
 \multicolumn{8}{l}{\hbox to 0pt{\parbox{180mm}{
 \footnotesize
 \footnotemark[$*$] HETE-2 trigger number.
 \par\noindent
 \footnotemark[$\dagger$] Photoelectric absorption with 90 \% confidence level uncertainties.
 \par\noindent
 \footnotemark[$\ddagger$] Blackbody temperature with 90 \% confidence level uncertainties.
 \par\noindent
 \footnotemark[$\S$] Emission radius with 90 \% confidence level uncertainties, where $d$ is the source distance. We used $d = 15$\,kpc.
 \par\noindent
 \footnotemark[$\|$] Spectral index with 90 \% confidence level uncertainties.
 \par\noindent
 \footnotemark[$\#$] X-ray (2-10\,keV) and gamma-ray (10-100\,keV) fluxes, 10$^{-8}$\,ergs\,cm$^{-2}$\,s$^{-1}$, with 68 \% confidence level uncertainties.
 }}}
 \endfoot
 \hline
 \multicolumn{8}{l}{\hbox to 0pt{\parbox{180mm}{
 \footnotesize
 \footnotemark[$*$] HETE-2 trigger number.
 \par\noindent
 \footnotemark[$\dagger$] Photoelectric absorption with 90 \% confidence level uncertainties.
 \par\noindent
 \footnotemark[$\ddagger$] Blackbody temperature with 90 \% confidence level uncertainties.
 \par\noindent
 \footnotemark[$\S$] Emission radius with 90 \% confidence level uncertainties, where $d$ is the source distance. We used $d = 15$\,kpc.
 \par\noindent
 \footnotemark[$\|$] Spectral index with 90 \% confidence level uncertainties.
 \par\noindent
 \footnotemark[$\#$] X-ray (2-10\,keV) and gamma-ray (10-100\,keV) fluxes, 10$^{-8}$\,ergs\,cm$^{-2}$\,s$^{-1}$, with 68 \% confidence level uncertainties.
 }}}
 \endlastfoot
 1566 & 16$_{-1.0}^{+12}$    & 5.4$_{-0.4}^{+0.5}$ & 19$_{-4}^{+4}$  & 2.1$_{-0.3}^{+0.3}$ & 45$_{-7}^{+7}$ & 202$_{-23}^{+23}$ & 10 (34) \\
 2770 & $<$24                & 4.5$_{-0.5}^{+0.7}$ & 18$_{-6}^{+5}$  & 1.6$_{-0.6}^{+0.5}$ & 23$_{-7}^{+7}$ & 93$_{-35}^{+35}$ & 11 (16) \\
 2771 & 18$_{-12}^{+19}$ & 3.8$_{-1.0}^{+1.8}$ & 18$_{-13}^{+13}$    & 2.1$_{-0.5}^{+0.4}$ & 20$_{-7}^{+7}$ & 63$_{-24}^{+24}$ & 24 (39) \\
 2773 & 56$_{-26}^{+56}$ & 7.6$_{-1.2}^{+1.7}$ & 10$_{-2}^{+5}$      & 2.7$_{-0.4}^{+0.6}$ & 42$_{-9}^{+9}$ & 254$_{-38}^{+38}$ & 84 (57) \\
 2800A & 21$_{-18}^{+46}$ & 5.3$_{-0.7}^{+1.3}$ & 11$_{-5}^{+4}$     & 2.0$_{-0.5}^{+0.6}$ & 15$_{-4}^{+4}$ & 79$_{-23}^{+23}$ & 18 (25) \\
 2800B & 34$_{-26}^{+57}$ & 5.3$_{-1.2}^{+2.3}$ & 9$_{-6}^{+7}$      & 2.6$_{-0.7}^{+1.3}$ & 14$_{-5}^{+5}$ & 50$_{-18}^{+18}$ & 22 (27) \\
 2802 & 24$_{-11}^{+19}$ & 7.3$_{-0.5}^{+0.6}$ & 14$_{-2}^{+2}$      & 2.0$_{-0.3}^{+0.3}$ & 37$_{-7}^{+7}$ & 338$_{-45}^{+45}$ & 55 (62) \\
 2806 & 27$_{-22}^{+32}$ & 6.3$_{-1.7}^{+2.0}$ & 9$_{-5}^{+8}$       & 2.6$_{-0.8}^{+1.1}$ & 23$_{-10}^{+10}$ & 86$_{-34}^{+34}$ & 13 (24) \\
\end{longtable}

\begin{longtable}{llllllll}
 \caption{Spectral parameters obtained using BB$+$PL for SGR\,1900$+$14.}\label{spc_list_bb_pow_2}
 \hline\hline
 Trigger \#\footnotemark[$*$] & $N_{\mathrm{H}}$\footnotemark[$\dagger$] &
 $kT_{\mathrm{BB}}$\footnotemark[$\ddagger$] & $R_{\mathrm{BB}}\left(\frac{d}{15\,{\mathrm{kpc}}}\right)$\footnotemark[$\S$] &
 $\alpha$\footnotemark[$\|$] & $F_{\mathrm{X}}$\footnotemark[$\#$] & $F_{\gamma}$\footnotemark[$\#$] & $\chi^2 (d.o.f.)$ \\
 & (10$^{22}$\,cm$^{-2}$) & (keV) & (km) & & & & \\
 \hline
 \endfirsthead
 \hline\hline
 Trigger \#\footnotemark[$*$] & $N_{\mathrm{H}}$\footnotemark[$\dagger$] &
 $kT_{\mathrm{BB}}$\footnotemark[$\ddagger$] & $R_{\mathrm{BB}}\left(\frac{d}{15\,{\mathrm{kpc}}}\right)$\footnotemark[$\S$] &
 $\alpha$\footnotemark[$\|$] & $F_{\mathrm{X}}$\footnotemark[$\#$] & $F_{\gamma}$\footnotemark[$\#$] & $\chi^2 (d.o.f.)$ \\
 & (10$^{22}$\,cm$^{-2}$) & (keV) & (km) & & & & \\
 \hline
 \endhead
 \hline
 \multicolumn{8}{l}{\hbox to 0pt{\parbox{180mm}{
 \footnotesize
 \footnotemark[$*$] HETE-2 trigger number.
 \par\noindent
 \footnotemark[$\dagger$] Photoelectric absorption with 90 \% confidence level uncertainties.
 \par\noindent
 \footnotemark[$\ddagger$] Blackbody temperature with 90 \% confidence level uncertainties.
 \par\noindent
 \footnotemark[$\S$] Emission radius with 90 \% confidence level uncertainties, where $d$ is the source distance. We used $d = 15$\,kpc.
 \par\noindent
 \footnotemark[$\|$] Spectral index with 90 \% confidence level uncertainties.
 \par\noindent
 \footnotemark[$\#$] X-ray (2-10\,keV) and gamma-ray (10-100\,keV) fluxes, 10$^{-8}$\,ergs\,cm$^{-2}$\,s$^{-1}$, with 68 \% confidence level uncertainties.
 }}}
 \endfoot
 \hline
 \multicolumn{8}{l}{\hbox to 0pt{\parbox{180mm}{
 \footnotesize
 \footnotemark[$*$] HETE-2 trigger number.
 \par\noindent
 \footnotemark[$\dagger$] Photoelectric absorption with 90 \% confidence level uncertainties.
 \par\noindent
 \footnotemark[$\ddagger$] Blackbody temperature with 90 \% confidence level uncertainties.
 \par\noindent
 \footnotemark[$\S$] Emission radius with 90 \% confidence level uncertainties, where $d$ is the source distance. We used $d = 15$\,kpc.
 \par\noindent
 \footnotemark[$\|$] Spectral index with 90 \% confidence level uncertainties.
 \par\noindent
 \footnotemark[$\#$] X-ray (2-10\,keV) and gamma-ray (10-100\,keV) fluxes, 10$^{-8}$\,ergs\,cm$^{-2}$\,s$^{-1}$, with 68 \% confidence level uncertainties.
 }}}
 \endlastfoot
 1568 & 43$_{-35}^{+77}$ & 7.0$_{-1.3}^{+1.3}$ & 7$_{-2}^{+4}$  & 2.6$_{-1.0}^{+1.9}$ & 26$_{-18}^{+18}$ & 157$_{-66}^{+66}$ & 20 (24) \\
 1571 & 15$_{-8}^{+12}$  & 6.6$_{-0.3}^{+0.3}$ & 13$_{-1}^{+2}$ & 2.2$_{-0.2}^{+0.3}$ & 65$_{-11}^{+11}$ & 419$_{-22}^{+22}$ & 71 (58) \\
 1572 & $<$145           & 6.8$_{-1.9}^{+1.2}$ & 6$_{-3}^{+4}$  & 2.1$_{-5.1}^{+2.9}$ & 21$_{-18}^{+18}$ & 153$_{-99}^{+99}$ & 7 (11) \\
 1578 & $<$48            & 6.8$_{-0.7}^{+0.8}$ & 7$_{-2}^{+2}$  & 2.8$_{-1.9}^{+6.3}$ & 24$_{-19}^{+19}$ & 112$_{-16}^{+16}$ & 5 (11) \\
 2310 & $<$9             & 3.7$_{-0.6}^{+0.6}$ & 15$_{-6}^{+5}$ & $<$2.0              & 16$_{-8}^{+8}$   & 52$_{-36}^{+36}$ & 8 (16) \\
\end{longtable}

%\appendix

%%%
% See the manual for the detail.
%%%


\clearpage

\begin{thebibliography}{}
\bibitem[Aptekar et al.(2001)]{apt01} Aptekar,~R.~L., Frederiks,~D.~D.,
   Golenetskii,~S.~V, Il'inskii,~V.~N., Mazets,~E.~P., \& Pal'shin,~V.~D. 2001, \apjs, 137, 227
\bibitem[Atteia et al.(2003)]{att03} Atteia,~J.~-L. \etal\ 2003, in Gamma-Ray Bursts
   and Afterglow Astronomy, ed. G. R. Ricker \& R. Vanderspek (Melville: AIP), 662, 17
\bibitem[Baring \& Harding(2001)]{bar01} Baring,~M.~G., \& Harding,~A.~K. 2001, \apj, 547, 929
\bibitem[Cea(2006)]{cea2006} Cea,~P. 2006, \aap, 450, 199
\bibitem[Cline et al.(1980)]{cli80} Cline,~T. \etal\ 1980, \apj, 237, L1
\bibitem[Cline et al.(2000)]{cli00} Cline,~T., Fredericks,~D.~D., Golenetskii,~S,
   Hurley,~K., Kouveliotou,~C., Mazets,~M., \& Van~Paradijs,~J. 2000, \apj, 531, 407
\bibitem[Colgate(1971)]{col71} Colgate,~S.~A. 1971, \apj, 163, 221
\bibitem[Cameron et al.(2005)]{cam05} Cameron,~P.~B. \etal\ 2005, \nat, 434, 1112
\bibitem[Corbel et al.(1997)]{cor97} Corbel,~S., Wallyn,~P., Dame,~T.~M., Durouchoux,~P.,
   Mahoney,~W.~A., Vilhu,~O., \& Grindlay,~J.~E. 1997, \apj, 478, 624
\bibitem[Corbel \& Eikenberry(2004)]{cor04} Corbel,~S., \& Eikenberry,~S.~S. 2004, \aap, 419, 191
\bibitem[Dennis(1985)]{den85} Dennis,~B.~R. 1985, \solphys, 100, 465
\bibitem[Dickey \& Lockman(1990)]{dic90} Dickey,~J.~M., \& Lockman,~F.~J. 1990, \araa, 28, 215
\bibitem[Duncan \& Thompson(1992)]{dun92} Duncan,~R., \& Thompson,~C. 1992, \apj, 392, L9
\bibitem[Duncan \& Thompson(1994)]{dun94} Duncan,~R., \& Thompson,~C. 1994,
   in Gamma-Ray Bursts. Am. Inst. Phys., ed. G.~J.~Fishman, J.~J.~Brainerd, \& K.~Hurley (Melville: AIP),
   307, 625
\bibitem[Duncan \& Thompson(2001)]{dun01} Duncan,~R., \& Thompson,~C. 2001, \apj, 561, 980
\bibitem[Evans et al.(1980)]{eva80} Evans,~W.~D. \etal\ 1980, \apj, 237, L7
\bibitem[Fenimore \& Laros(1994)]{fen94} Fenimore,~E.~E., \& Laros,~J.~G. 1994, \apj, 432, 742
\bibitem[Fenimore, Klebesadel \& Laros(1996)]{fen96} Fenimore,~E.~E., Klebesadel,~R.~W., \&
   Laros,~J.~G. 1996, \apj, 460, 964
\bibitem[Feroci et al.(2001)]{fer01} Feroci,~M., Hurley,~K., Duncan,~R.~C., \& Thompson,~C. 2001,
   \apj, 549, 1021
\bibitem[Feroci et al.(2004)]{fer04} Feroci,~M., Calliandro,~G.~A., Massaro,~E.,
   Mereghetti,~S., \& Woods,~P.~M. 2004, \apj, 612, 408
\bibitem[Gaensler et al.(2001)]{gae01} Gaensler,~B., Slane,~P., Gotthelf,~E., \& Vasisht,~G. 2001,
   \apj, 559, 963
\bibitem[G\"{o}\u{g}\"{u}\c{s} et al.(2000)]{gog00} G\"{o}\u{g}\"{u}\c{s},~E., Woods,~P.~M.,
   Kouveliotou,~C., Paradijs,~J.~V., Briggs,~M.~S., Duncan,~R.~C., \& Thompson,~C. 2000, \apj, 532, L121
\bibitem[G\"{o}\u{g}\"{u}\c{s} et al.(2001)]{gog01} G\"{o}\u{g}\"{u}\c{s},~E., Kouveliotou,~C,
   Woods,~P.~M., Thompson,~C., Duncan,~R.~C., \& Briggs,~M.~S. 2001, \apj, 558, 228
\bibitem[Golenetskii et al.(1987)]{gol87} Golenetskii,~S.~V., Aptekar,~R.~L., Guryan,~Y.~A.,
   Il'inskii,~V.~N., \& Mazets,~E.~P. 1987, Sov. Astron. Lett., 13, 166
\bibitem[G\"{o}tz et al.(2004)]{got04} G\"{o}tz,~D., Mereghetti,~S., Mirabel,~I.~F.,
   \& Hurley,~K. 2004, \aap, 417, L45
\bibitem[G\"{o}tz et al.(2006)]{got06} G\"{o}tz,~D. \etal\ 2006, \aap, 445, 313
\bibitem[Heger et al.(2003)]{heg03} Heger,~A., Fryer,~C.~L., Woosley,~S.~E.,
   Langer,~N., \& Hartmann,~D.~H. 2003, \apj, 591, 288
\bibitem[Holland et al.(2005)]{hol05} Holland,~S. T., Barthelmy,~S., Beardmore,~A.,
   Gehrels,~N., Kennea,~J., Page,~K., Palmer,~D., \& Rosen,~S. 2005, GCN Circ., 4034
\bibitem[Hurley et al.(1994)]{hur94} Hurley,~K., Sommer,~K., Kouveliotou,~C.,
   Fishman,~G., Meegan,~C., Cline,~T., Boer,~M., \& Niel,~M., 1994, \apj, 431, L31
\bibitem[Hurley et al.(1996)]{hur96} Hurley,~K., \etal\ 1996, \apj, 463, L13
\bibitem[Hurley et al.(1999a)]{hur99a} Hurley,~K., \etal\ 1999, \nat, 397, 41
\bibitem[Hurley et al.(1999b)]{hur99b} Hurley,~K., Kouveliotou,~C., Woods,~P., Cline,~T.,
   Butterworth,~P., Mazets,~E., Golenetskii,~S., \& Frederics,~D. 1999, \apj, 510, L107
\bibitem[Hurley et al.(1999c)]{hur99c} Hurley,~K., \etal\ 1999, \apj, 510, L111
\bibitem[Hurley et al.(2005)]{hur05} Hurley,~K., \etal\ 2005, \nat, 434, 1098
\bibitem[Ibrahim et al.(2001)]{ibr01} Ibrahim,~A.~I., \etal\ 2001, \apj, 558, 237
\bibitem[Kagan(1999)]{kag99} Kagan,~Y.~Y. 1999, Pure Appl. Geophys., 155, 537
\bibitem[Kohri \& Yamada(2002)]{koh02} Kohri,~K., \& Yamada,~S., 2002, \prd, 65, 043006
\bibitem[Kouveliotou et al.(1987)]{kou87} Kouveliotou,~C., \etal\ 1987, \apj, 322, L21
\bibitem[Kouveliotou et al.(1998)]{kou98} Kouveliotou,~C., \etal\ 1998, \nat, 393, 235
\bibitem[Kouveliotou et al.(1999)]{kou99} Kouveliotou,~C., \etal\ 1999, \apj, 510, L115
\bibitem[Lamb et al.(2003)]{lam03} Lamb,~D.~Q., \etal\ 2003, GCN Circ., 2351
\bibitem[Laros et al.(1986)]{lar86} Laros,~J.~G., Fenimore,~E.~E., Fikani,~M.~M., Klebesadel,~R.~W.,
   \& Barat,~C. 1986, \nat, 322, 152
\bibitem[Lyutikov(2003)]{lyu03} Lyutikov,~M. 2003, \mnras, 346, 540
\bibitem[Mazets et al.(1979b)]{maz79a} Mazets,~E.~P., Golenetskii,~S.~V., Il'inskii,~V.~N.,
   Aptekar,~R.~L., \& Gur'yan,~Yu.~A. 1979, \nat, 282, 587
\bibitem[Mazets, Golenetskii \& Gur'yan(1979a)]{maz79b} Mazets,~E.~P., Golenetskii,~S.~V.,
    \& Gur'yan,~Yu.~A. 1979, Soviet Astron. Lett., 5, 343
\bibitem[Mazets et al.(1981)]{maz81} Mazets,~E.~P., \etal\ 1981, \apss, 80, 3
\bibitem[McClure-Griffiths \& Gaensler(2005)]{mcc05} McClure-Griffiths,~N.~M., \&
   Gaensler,~B.~M. 2005, \apj, 630, L161
\bibitem[Michel \& Dessler(1981)]{mic81} Michel,~F.~C. \& Dessler,~A.~J. 1981, \apj, 251, 654
\bibitem[Nakagawa et al.(2005)]{nak05} Nakagawa,~Y.~E. \etal\ 2005, Il Nuovo Cimento C, vol. 28, Issue 4, p.625
\bibitem[Norris(2002)]{nor02} Norris,~J.~P. 2002, \apj, 579, 386
\bibitem[Norris \& Bonnell(2006)]{nor06} Norris,~J.~P. \& Bonnell,~J.~T. 2006, \apj, 643, 266
\bibitem[Olive et al.(2004)]{oli04} Olive,~J.~-F., \etal\ 2004, \apj, 616, 1148
\bibitem[Paczy\'{n}ski(1992)]{pac92} Paczy\'{n}ski,~B. 1992, Acta Astron., 42, 145
\bibitem[Palmer et al.(2005)]{pal05} Palmar,~D.~M., \etal\ 2005, \nat, 434, 1107
\bibitem[Perna, Hernquist \& Narayan(2000)]{per00} Perna,~R., Hernquist,~L., \& Narayan,~R. 2000, \apj, 541, 344
\bibitem[Ricker et al.(2003)]{ric03} Ricker,~G., \etal\ 2003, in Gamma-Ray Bursts
   and Afterglow Astronomy, ed. G.~R.~Ricker \& R.~Vanderspek (Melville: AIP), 662, 3
\bibitem[Rybicki \& Lightman (1979)]{ryb79} Rybicki,~G.~B., \& Lightman,~A.~P. 1979,
   Radiative Processes in Astrophysics, John Wiley \& Sons, Inc, New York
\bibitem[Shirasaki et al.(2003)]{shi03} Shirasaki,~Y., \etal\ 2003, \pasj, 55, 1033
\bibitem[Sonobe et al.(1994)]{son94} Sonobe,~T., Murakami,~T., Kulkarni,~S.~R., Aoki,~T.,
   \& Yoshida,~A. 1994, \apj, 436, L23
\bibitem[Strohmayer \& Ibrahim(1998)]{str98} Strohmayer,~T.~E., \& Ibrahim,~A. 1998,
   in Fourth Huntsville Symp. on Gamma-Ray Bursts, ed. C.~A.~Meegan, R.~D.~Preece,
   \&     T.~M.~Koshut (Woodbury: AIP), AIP Conf. Proc., 428, 947
\bibitem[Terasawa et al.(2005)]{ter05} Terasawa,~T., \etal\ 2005, \nat, 434, 1110
\bibitem[Thompson \& Duncan(1995)]{tho95} Thompson,~C., \& Duncan,~R. 1995, \mnras, 275, 255
\bibitem[Thompson \& Duncan(1996)]{tho96} Thompson,~C., \& Duncan,~R. 1996, \apj, 473, 322
\bibitem[Vasisht et al.(1994)]{vas94} Vasisht,~G., Kulkarni,~S., Frail,~D.,
   \& Greiner,~J. 1994, \apj, 431, L35
\bibitem[Vrba et al.(2000)]{vrb00} Vrba,~F.~J., Henden,~A.~A., Luginbuhl,~C.~B., Guetter,~H.~H.,
   Hartmann,~D.~H., \& Klose,~S. 2000, \apj, 533 L17
\bibitem[Wang, Chakrabarty \& Kaplan(2006)]{wan06} Wang,~Z., Chakrabarty,~D, \& Kaplan,~L. 2006, \nat, 440, 772
\end{thebibliography}
\end{document}